\documentclass[twocolumn,oldversion,a4paper]{aa}
\usepackage{graphicx}
\usepackage{natbib}
\usepackage{txfonts}
\usepackage{float}
\usepackage{pdflscape}
\bibpunct{(}{)}{;}{a}{}{,} 

\def\lsim{\mathrel{\rlap{\lower4pt\hbox{\hskip1pt$\sim$}}
    \raise1pt\hbox{$<$}}}                
\def\gsim{\mathrel{\rlap{\lower4pt\hbox{\hskip1pt$\sim$}}
    \raise1pt\hbox{$>$}}}  

\begin{document} 

\title{The implications of the surprising existence of a large, massive
CO disk in a distant protocluster}

\titlerunning{Large, massive molecular disk at z$\sim$2}
\author{H.~Dannerbauer\inst{1,2,3}, M.~D.~Lehnert\inst{4}, B.~Emonts\inst{5}, B.~Ziegler\inst{3}, B.~ Altieri\inst{6}, C. De Breuck\inst{7}, N.~Hatch\inst{8}, T.~Kodama\inst{9}, Y.~Koyama\inst{9,10}, J.~D.~Kurk\inst{11}, T.~Matiz\inst{3}, G.~Miley\inst{12}, D.~Narayanan \inst{13}, R.~P.~Norris\inst{14,15}, R.~Overzier\inst{16}, H.~J.~A.~R\"ottgering\inst{12}, M.~Sargent\inst{17}, N.~Seymour\inst{18}, M.~Tanaka\inst{9}, I.~Valtchanov\inst{6}, and D.~Wylezalek\inst{19}}

\authorrunning{Dannerbauer, Lehnert, Emonts et al.}
\institute{Instituto de Astrof\'{i}sica de Canarias (IAC), E-38205 La Laguna, Tenerife, Spain\\\email{helmut@iac.es}
\and
Universidad de La Laguna, Dpto. Astrof\'{i}sica, E-38206 La Laguna, Tenerife, Spain
\and
Universit\"at Wien, Institut f\"ur Astrophysik, T\"urkenschanzstra\ss e 17, 1180 Vienna, Austria
\and
Sorbonne Universit\'es, UPMC Univ Paris 6 et CNRS, UMR 7095, Institut d'Astrophysique de Paris, 98 bis Bd Arago, 75014 Paris, France
\and 
Centro de Astrobiolog\'{i}a (INTA-CSIC), Ctra de Torrej\'{o}n a Ajalvir, km 4, E-28850 Torrej\'{o}n de Ardoz, Madrid, Spain
\and
\textit{Herschel} Science Centre, European Space Astronomy Centre, ESA, 28691 Villanueva de la Ca\~{n}ada, Spain
\and
European Southern Observatory, Karl Schwarzschild Stra\ss e 2, 85748 Garching, Germany 
\and
School of Physics and Astronomy, University of Nottingham, University Park, Nottingham NG7 2RD, UK
\and Optical and Infrared Astronomy Division, National Astronomical Observatory of Japan, Mitaka, Tokyo 181-8588, Japan
\and Department of Space Astronomy and Astrophysics, Institute of Space and Astronautical Science (ISAS)
Japan Aerospace Exploration Agency (JAXA), 3-1-1 Yoshinodai, Chuo-ku, Sagamihara, Kanagawa 252-5210, Japan 
\and 
Max-Planck-Institut f\"ur extraterrestrische Physik, Giessenbachstra\ss e 1, 85748 Garching, Germany 
\and
Leiden Observatory, PO Box 9513, 2300 RA Leiden, the Netherlands
\and 
University of Florida Department of Astronomy, 211 Bryant Space Sciences Center, Gainesville, FL, USA
\and
CSIRO Astronomy \& Space Science, PO Box 76, Epping, NSW 1710, Australia
\and
Western Sydney University, Locked Bag 1797, Penrith South, NSW 1797, Australia
\and
Observat\'{o}rio Nacional, Rua Jos\'{e} Cristino, 77. CEP 20921-400, S\~{a}o Crist\'{o}v\~{a}o, Rio de Janeiro-RJ, Brazil
\and
Astronomy Centre, Department of Physics and Astronomy, University of Sussex, Brighton BN1 9QH, UK
\and
International Center for Radio Astronomy Research, Curtin University, GPO Box U1987, Perth, WA 6845, Australia
\and
Johns Hopkins University Bloomberg Center -- Department of Physics \& Astronomy, 3400 N. Charles Street, Baltimore, MD, 21218, USA
}

\date{Received/accepted}

\abstract{It is not yet known if the properties of molecular gas in
distant protocluster galaxies are significantly affected by their
environment as galaxies are in local clusters. Through a deep, 64
hours of effective on-source integration with the Australian Telescope
Compact Array (ATCA), we discovered a massive, M$_\mathrm{mol}$ =
$2.0\pm0.2$~$\times$~10$^{11}$~M$_{\sun}$, extended, $\sim$40 kpc,
CO(1-0)-emitting disk in the protocluster surrounding the radio galaxy,
MRC\,1138$-$262. The galaxy, at $z_{CO}=2.1478$, is a clumpy,
massive disk galaxy, M$_{\ast}\sim5\times10^{11}$~M$_{\sun}$, which
lies 250~kpc in projection from MRC\,1138$-$262 and is a known H$\alpha$
emitter, HAE229. HAE229 has a molecular gas fraction of $\sim$30\%. The
CO emission has a kinematic gradient along its major axis, centered on
the highest surface brightness rest-frame optical emission, consistent
with HAE229 being a rotating disk. Surprisingly, a significant fraction
of the CO emission lies outside of the UV/optical emission. In
spite of this, HAE229 follows the same relation between star-formation
rate and molecular gas mass as normal field galaxies.

HAE229 is the first CO(1-0) detection of an ordinary, star-forming galaxy
in a protocluster. We compare a sample of cluster members at $z>0.4$
that are detected in low-order CO transitions with a similar sample of
sources drawn from the field. We confirm findings that the CO-luminosity
and FWHM are correlated in starbursts and show that this relation is valid
for normal high-z galaxies as well as those in overdensities. We do not
find a clear dichotomy in the integrated Schmidt-Kennicutt relation for
protocluster and field galaxies. Our results suggest that environment
does not impact the ``star-formation efficiency'' or the molecular
gas content of high-redshift galaxies. Not finding any environmental
dependence in these characteristics, especially for such an extended
CO disk, suggests that environmentally-specific processes such as ram
pressure stripping are not operating efficiently in (proto)clusters. We
discuss why this might be so.}


\maketitle

\section{Introduction}
\label{sec:intro}

Over the last decade, detections of molecular line emission in
high-redshift galaxies have become routine \citep[e.g.,][]{car13}. These
detections mainly came from observations of the bright high order
transitions of CO and generally from extreme source populations
such as submillimeter galaxies \citep[SMGs; see ][for detailed
reviews]{bla02,cas14} or high-z QSOs \citep[e.g.,][]{wal04} which are
intrinsically gas-rich (M$_{\rm H_2}\sim$~few times$~10^{10}$~M$_{\sun}$)
or from strongly lensed sources \citep[e.g.,][]{bak04, les10, wei13,
spi14, can15, dye15, sw15, spi15, ara16, bet16, har16, sha16}. Very
few detections are available for normal star-forming galaxies
\citep[e.g.,][]{dad08,dad10a,dad15,dan09,tac10,tac13,gen15}. These are
found to have substantial reservoirs of molecular gas, but are converting
their gas into stars with a lower efficiency.

The number of CO detections of galaxies which lie in
overdensities (protoclusters) at $z>1$ is still small
\citep[e.g.,][]{dad09a,cap11,ara12,wag12,cas16,wan16}. This is
unfortunate, the limited sample size of well-studied protocluster
galaxies in CO hampers our ability to study the build-up of
galaxy populations in clusters.  Almost all of the CO detections
of protocluster galaxies are in the environments of high-redshift
radio galaxies \citep{ivi08,ivi12,cas13,emo14}. \citet{tad14} detected
three galaxies -- two robustly and one tentatively -- in CO(1-0) in the
protocluster surrounding USS~1558$-$003 at $z=2.53$. These three galaxies
were all originally identified as H$\alpha$ emitters (HAEs). They conclude
that these HAEs, based on their estimated star-formation efficiencies,
SFE=SFR/M$_\mathrm{mol}$ (the ratio of star-formation rate and molecular
gas mass), are gas-rich major mergers. \citet{gea11} and \citet{ara12}
find that the SFE of IR-bright cluster members are similar to disk-like
galaxies at lower redshifts. On the contrary, \citet{ivi13} find that two
of four discovered CO-bright galaxies within a region of $\sim$100~kpc
do have high SFEs. \citet{jab13} made the first detailed study of
how molecular gas properties depend on the environment beyond the
local universe.  They find that, at intermediate redshifts, $z\sim0.4$,
environment is starting to affect the cold gas content of the most massive
galaxies in clusters.  Since we know that cluster galaxies in the nearby
universe and at moderate redshifts are relatively gas poor \citep{cha80,
cha86, vol01a, jab13}, the crucial question to investigate is: When
does environment begin to play a significant role in shaping the gas
content of galaxies \citep{vol01b, gne03a, gne03b,hus16}?  Of course,
like many questions in astrophysics, answering this is hampered by small
sample sizes and the lack of systematic studies \citep[cf.][]{cha15}.

In one of the best studied protocluster fields, MRC\,1138$-$262 at $z=2.16$
\citep[e.g.,][]{kur00,kur04a,kur04b,pen00,hat11,dan14,emo16}, using
the Australia Telescope Compact Array (ATCA), \citet{emo13} tentatively
detected the CO(1-0) emission from the H$\alpha$ bright galaxy, HAE229
\citep{kur04b,doh10}. \citet{dan14} found evidence that HAE229 -- about
30$^{\prime\prime}$ ($\sim$250~kpc in projected distance) from the radio
galaxy MRC\,1138$-$262 -- is an SMG and part of the overdensity of dusty
starbursts in the field of MRC\,1138$-$262. Its redshift also indicates
it is a protocluster member \citep{kui11}. Here we present a robust
detection of the CO(1-0) line of this source adding new, deeper and
higher-resolution ATCA data \citep{emo13}.  Other than the radio galaxy
MRC\,1138$-$262 \citep{emo13,emo16}, HAE229 is the first unambiguously
confirmed gas-rich member of this well-studied protocluster.

The structure of this paper is as follows. Sections 2 and 3 describe the
properties of HAE229 and our new ATCA observations. In
Sect.~4 we present the results of the CO(1-0) observations of HAE229
and in Sect.~5, we discuss the properties of this gas-rich HAE and
compare the molecular gas properties of protocluster galaxies and
field galaxies. We adopt the cosmological parameters  $\Omega_{\rm
m}=0.27$, $\Omega_{\Lambda}=0.73$, and H$_{0}=71$~km~s$^{-1}$~Mpc$^{-1}$
\citep{spe03,spe07}. At redshift $z=2.16$, 1 arcsec corresponds to
8.4~kpc. All magnitudes in this paper are on the AB magnitude scale
\citep{oke83} and we assume a Salpeter IMF \citep{sal55} in our analysis.

\section{A brief history of HAE229}
\label{sec:hae229}

\citet{kur04a} discovered HAE229 through H$\alpha$ narrow-band
imaging at the approximate redshift of the radio galaxy. The
excess emission in the narrow band image of HAE229 was subsequently
confirmed spectroscopically to be H$\alpha$ emission at $z=2.1489$
\citep[][see also \citealt{koy13}]{kur04b}. \citet{doh10} found
that HAE229 (\#464 in their paper) is a massive, dust-obscured
star-forming red galaxy, [J--K$]>2.41$, with a stellar mass,
M$_{\ast}=5.1^{+1.5}_{-2.0}~\times~10^{11}$~M$_{\sun}$, and a
star-formation rate estimated from spectral energy distribution
(SED) fitting, SFR$_\mathrm{SED}=35\pm6$~M$_{\sun}$ yr$^{-1}$.
\citet{ogl12} observed this source with {\it Spitzer IRS},
detecting PAH emission at $7.7~\mu$m, concluding that the star
formation in HAE229 is heavily obscured and implying a much higher
SFR,   SFR$_\mathrm{PAH}=880$~M$_{\sun}$ yr$^{-1}$, than was estimated
previously \citep[e.g.,][]{doh10}.  HAE229 is one of the most massive
HAEs embedded in the large scale structure at $z=2.16$ \citep[see Fig.~6
in][and Fig.~8 in \citealt{dan14}]{koy13}.  Finally, when observing
the radio galaxy MRC\,1138$-$262 with ATCA, \citet{emo13} serendipitously
found a tentative CO(1-0) emission line at $z_\mathrm{CO}=2.147$ at the
position of HAE229 and \citet{dan14} associated this source with a SCUBA
submm detection \citep{ste03}.

\section{Observations}

\subsection{ATCA observations of CO(1-0)}\label{subsec:atcaobs}

Our CO(1-0) observations of HAE229 were performed with the Australia
Telescope Compact Array during April 2011 - Feb 2015 in the H75,
H168, H214, 750A, 750D and 1.5A array configurations and only
including baselines ranging $31-800$~m in our reduction. Data from
the longest baselines of the 1.5A array configuration were discarded
because these data were obtained during day-time under moderate
weather conditions. Excluding them from the reduction resulted
in a more uniform uv-coverage and a more robust image. Our total
on-source integration time is $\sim$90~hours.  Our primary goal
with these observations was to obtain ultra-deep data on the radio
galaxy within the protocluster, MRC\,1138$-$262. Hence the pointing
center of the observations was 30 arcsec east of our target HAE\,229
\citep[see][]{emo13}. This increased the effective noise at its location
by a factor of $\sim$1.4 as a result of the primary beam correction
(FWHM$_\mathrm{PrimBeam}$\,$\sim$\,77\,arcsec). The effective integration
time is therefore only 60 hours or 2/3$^{rd}$ of the total on-source time.
Observations were centered around 36.5~GHz, using a channel width of 1
MHz and an effective bandwidth of 2~GHz.

\begin{figure}[ht]
\centering
\includegraphics[width=8cm,angle=0]{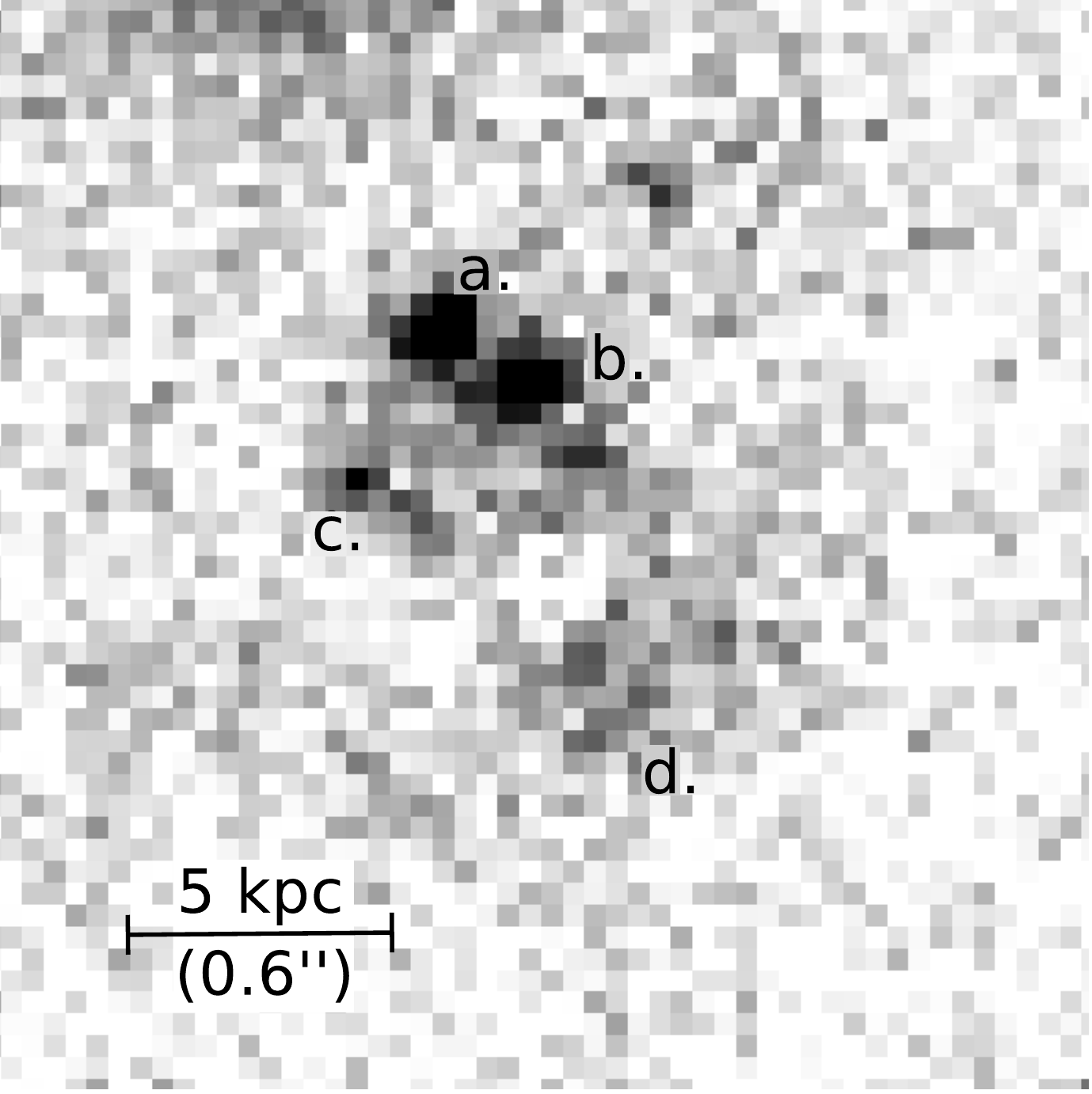}
\caption{$3^{\prime\prime}\times 3^{\prime\prime}$ \textit{HST} F814W image of the
clumpy galaxy, HAE229. The size of HAE229 is $1\farcs2\times0\farcs6$
(10$\times$5~kpc). The regions a and b are clumpy and knot-like whereas
areas c and d have more diffuse rest-frame UV emission morphology.}
\label{fig:hstmorph}
\end{figure}

Phase and bandpass calibration were performed by observing the strong
calibrator PKS\,1124$-$186 every 5\,$-$\,12 min. The frequency of the
observations depended on the weather conditions. However, given the
8.2$^{\circ}$ distance of PKS\,1124-186 from our target, we used the
weaker but closer (2.8$^{\circ}$) calibrator PKS\,1143-287 for phase
calibration in the more extended 750A/D and 1.5A array configurations.
The bandpass calibration scans of PKS\,1124-186 were taken approximately
every hour. An absolute flux scale was determined using observations
of Mars for array configurations, H75 and H168, the ultra-compact H\,II region G309 for H214 and PKS\,1934-638 for array
configurations, 750A/D and 1.5A \citep{emo11mnras}.\footnote{The reason for
altering the flux calibrators is that the reliability of the absolute
flux scale of PKS\,1934-638 was still
questionable during the epoch 2011-2013. Mars was not always visible
and G309 \citep[with flux bootstrapped from Uranus;][]{emo11mnras}
is fully resolved in the longer-baseline 750A/D and 1.5A array
configurations.} The strong radio continuum of MRC\,1138$-$262
allowed us to verify that the flux scaling between all observations
stayed within the typical 30$\%$ accuracy for flux calibration at the
ATCA. These data were reduced in MIRIAD \citep{sau95} and analysed with
the KARMA software \citep{goo96}, following the strategy described in
\citet[][]{emo13}. The continuum-subtracted line-data products that we
present in this paper were weighted using a robustness parameter of +1
\citep{bri95}, binned into 34 km\,s$^{-1}$ channels and subsequently
Hanning smoothed to an effective velocity resolution of 68 km\,s$^{-1}$
(equivalent to two 34 km\,s$^{-1}$ binned channels). This procedure results
in a root-mean-square noise of 0.12 mJy\,beam$^{-1}$ per channel in the
region of HAE229, after correcting for the primary beam response. At
the half-power point, the synthesized beam is $4\farcs7\times\,3\farcs1$
(PA~$=-6^{\circ}$). Velocities in this paper are defined in the optical
barycentric reference frame with respect to $z$\,=\,2.1478.

\subsection{HAWK-I}\label{subsec:hawkiobs}

We mapped the field around the radio galaxy, MRC\,1138$-$262, with the
near-infrared wide-field imager HAWK-I at the ESO/VLT. The observations
were taken in the Y-, H-, and K$_{s}$-bands during February/March 2012,
April/May/July 2013 and January/February 2015 in service mode. The seeing
was $0\farcs4-0\farcs6$ during these observations. The dithered HAWK-I data
were reduced using the ESO/MVM data reduction pipeline \citep{vda04},
following the standard reduction steps for near-infrared imaging data.

\begin{figure*}[!h]
\centering
\includegraphics[width=10cm,angle=0]{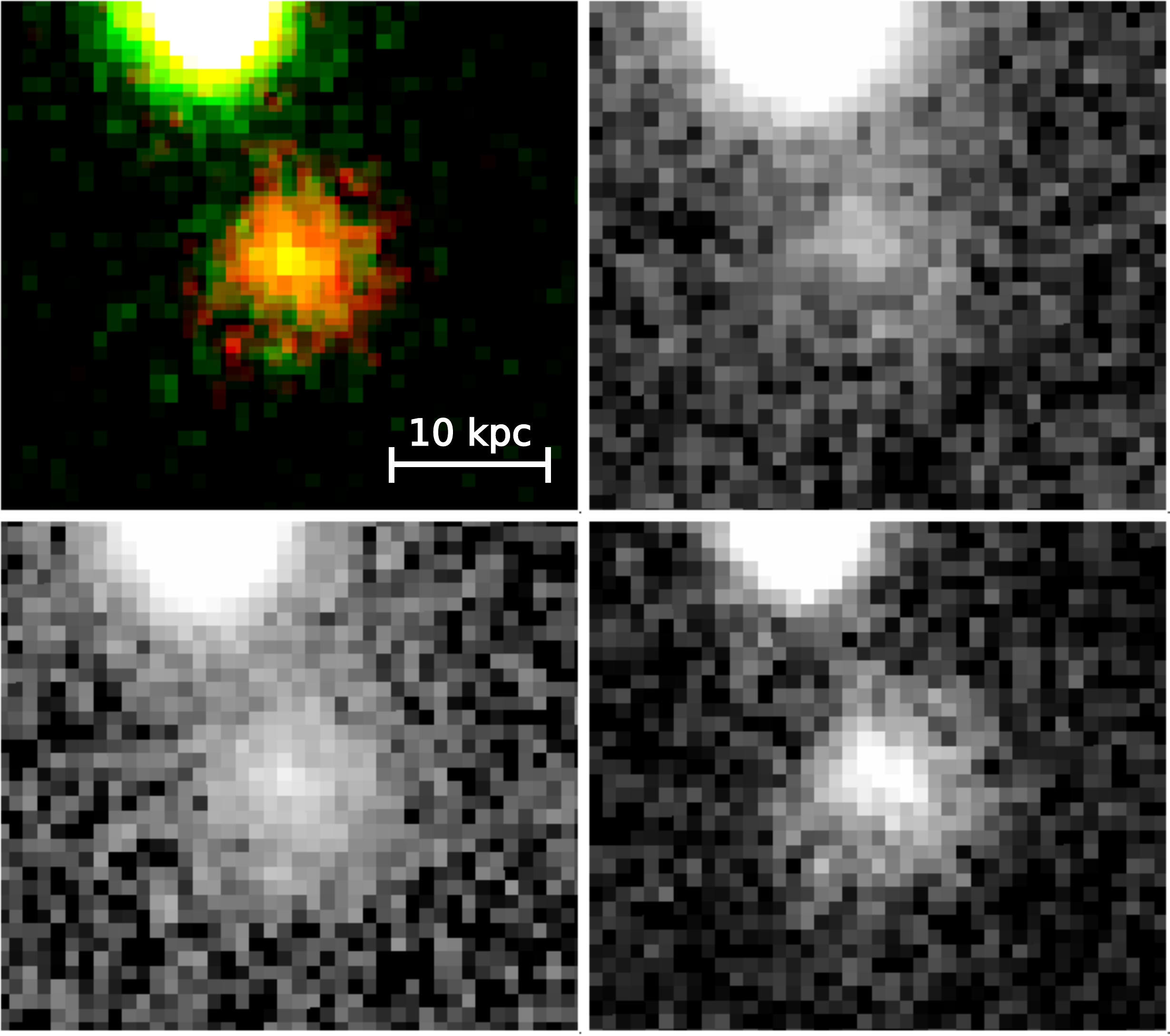}
\caption{$4.2^{\prime\prime}\times 3.7^{\prime\prime}$ regions centered
on HAE229. \textit{Top left panel:} A composite 3-color image made
by combining the Y-, H-, and K-band color images taken with HAWK-I on
the ESO/VLT (see text and Table~\ref{tab:flux} for details). To show
the rest-frame blue and red optical morphology, we show each color
components of the composite separately, namely, Y-band (approximately
rest-frame U-band, \textit{top right}), H-band (approximately rest-frame
B- or V-band, \textit{bottom left}) and  K$_{s}$-band (approximately
rest-frame R-band, \textit{bottom right}). The Y-band image has a
morphology similar to that of the F814W image and no offset is
detected between the rest-frame UV and rest-frame optical regions. Note
that with increasing wavelength, the emission becomes less clumpy, more
regular and a red nucleus is evident. The dynamical center of the CO(1-0)
emission corresponds to the highest surface brightness region of the H-
and K$_{s}$-band images suggesting this region is also the center of
mass of the galaxy. The bright galaxy above HAE229 in all panels lies
at $z_{phot}\approx0.5$ \citep{tan10}.}
\label{fig:gallery}
\end{figure*}

\section{Results}
\label{sec:results}

HAE229 has been observed with \textit{HST}/ACS through the F475W and F814W
filters \citep{mil06} and with \textit{HST}/NICMOS through the $J_{110}$ and
H$_{160}$ filters \citep{zir08}. In all of these images, we only detect
HAE229 in the ACS F814W band (rest-frame $\sim$$2560\AA$)\footnote{We
note that \citet{koy13} discuss the \textit{HST} F814W imaging of 54 HAEs,
one of them is HAE229.}. There are several regions of higher surface
brightness rest-frame UV emission seen in the \textit{HST} image (labeled a, b,
c, d; Fig.~\ref{fig:hstmorph}) embedded in more diffuse, lower surface
brightness emission. Two of them, a and b, are ``clumps'' of UV emission
(meaning marginally extended, point-like sources), while c and d have
fainter clumps superposed on diffuse continuum emission. The size of
this clumpy galaxy --- with marginally resolved sources evident in
the \textit{HST}/ACS light distribution system --- is $1\farcs2\times0\farcs6$
(10$\times$5~kpc).  Despite its clumpy structure, estimating the
``Gini'' coefficient \citep[e.g.,][]{abr03} of the {\it HST }F814W image of
HAE229 suggests that its light is dominated by a uniform, lower surface
brightness diffuse component \citep[see ][ for details]{koy13}.  We detect
this source in Subaru MOIRCS and VLT HAWK-I near-infrared imagery but
not in shallower {\it HST }NIR-images. At near-infrared wavelengths, HAE229
becomes much more regular in appearance (Fig.~\ref{fig:gallery}).  We do
not see an offset of centers between the rest-frame UV ($\sim$$2560~\AA$)
and rest-frame optical regions ($\sim$$3063- 7300~\AA$) of HAE229. The
comparable images of normal SFGs (star-forming galaxies) shows that they
often consist of clumps within diffuse continuum emission similar to what
we observe for HAE229 \citep[see the image montages in, e.g.,][]{tac13}.
In addition, the highest surface brightness region in the near-infrared,
presumably the center of mass of the stellar component, is very red
and is responsible for giving HAE229 its overall red color.  Again,
this is found among many distant disk galaxies, especially ones that
are similarly massive \citep[e.g.,][]{pan09}

\begin{figure}[th] 
\centering
\includegraphics[width=9cm,angle=0]{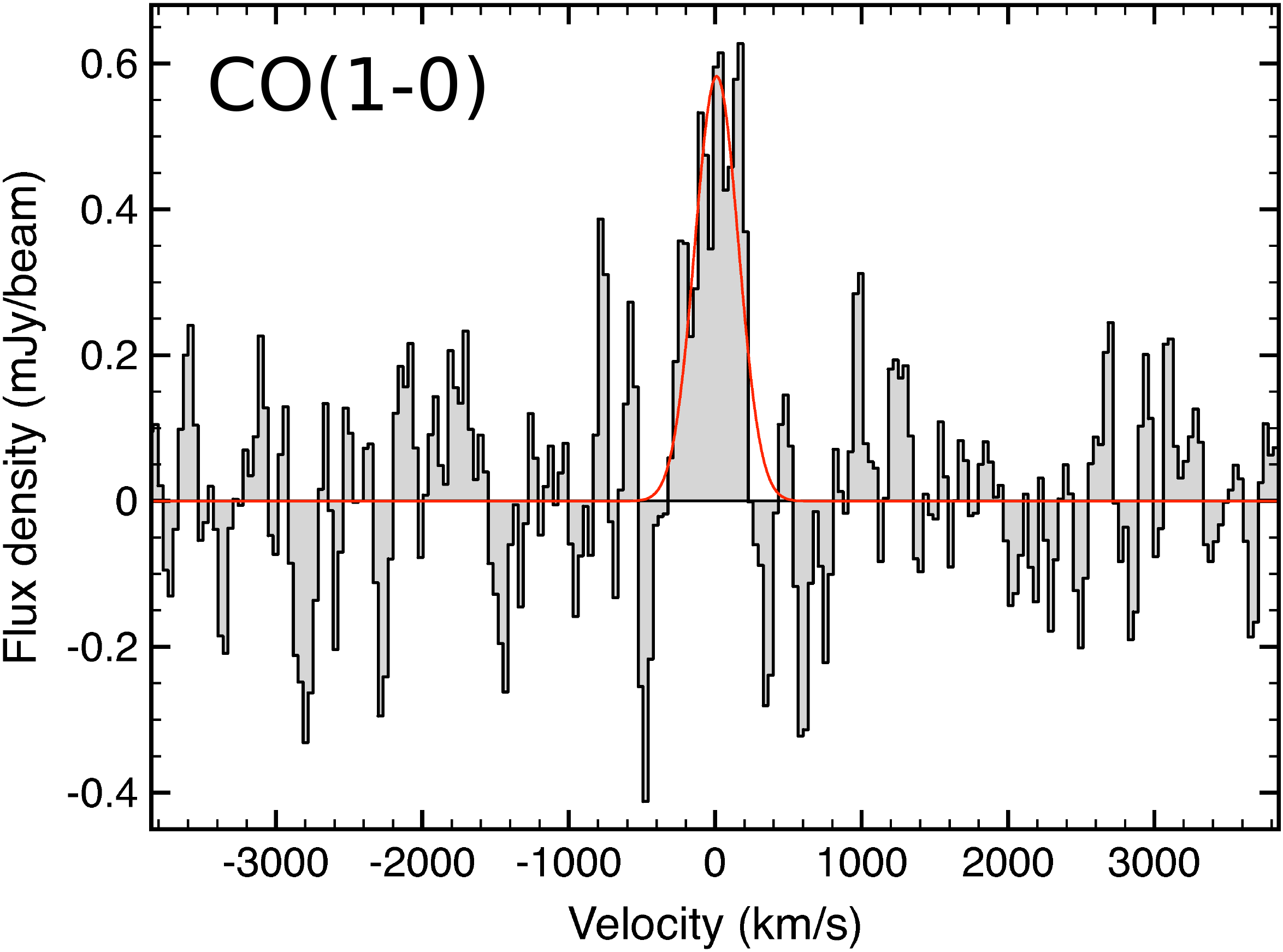}
\caption{CO(1-0) spectrum of HAE229 from data taken with the
compact hybrid ATCA configurations (for which the signal is spatially
unresolved). The red line shows a Gaussian fit from which we derived $z$,
$L’_{\rm CO(1-0)}$ and FWHM$_{\rm CO(1-0)}$.}
\label{fig:co1dspec}
\end{figure}

The CO(1-0) transition is now robustly detected in
our new data for HAE229, with a total significance of
$\sim$7$\sigma$ (Fig.~\ref{fig:co1dspec}) We measure
a peak flux, $S_{\nu}=0.57\pm0.06$~mJy beam$^{-1}$,
at the position: RA$_\mathrm{2000.0}$~=~11:40:46.05 and
Dec$_\mathrm{2000.0}$~=~$-$26:29:11.2s (Table~\ref{tab:position}). The
full-width at half maximum (FWHM) of the line is $359\pm34$~km s$^{-1}$
and we obtain an integrated flux, $I_\mathrm{CO(1-0)}=0.22\pm0.03$~Jy km
s$^{-1}$. The CO line redshift $z_\mathrm{CO(1-0)}=2.1478\pm0.0002$
agrees with the redshift estimate obtained using H$\alpha$
\citep{kur04b, doh10}. Both the measured redshift and line FWHM are
consistent with the results of \citet{emo13} for their tentative
detection.  However, the flux density increased by 80\%.  We derive
$L^{'}_\mathrm{CO(1-0)}=5.0\pm0.07\times10^{10}$~K~km~s$^{-1}$~pc$^{2}$
from the current data set. We stress that the uncertainties in
the flux and luminosity estimates are measurement errors, and do
not not include the 30\% uncertainty in absolute flux calibration
(Sect.~\ref{subsec:atcaobs}), or potential errors in the primary beam
calibration (which may be important considering that HAE229 is located
close to the edge of the primary beam; Sect.~\ref{subsec:atcaobs}).

\begin{figure*}[t]
\begin{centering}
\includegraphics[width=16cm,angle=0]{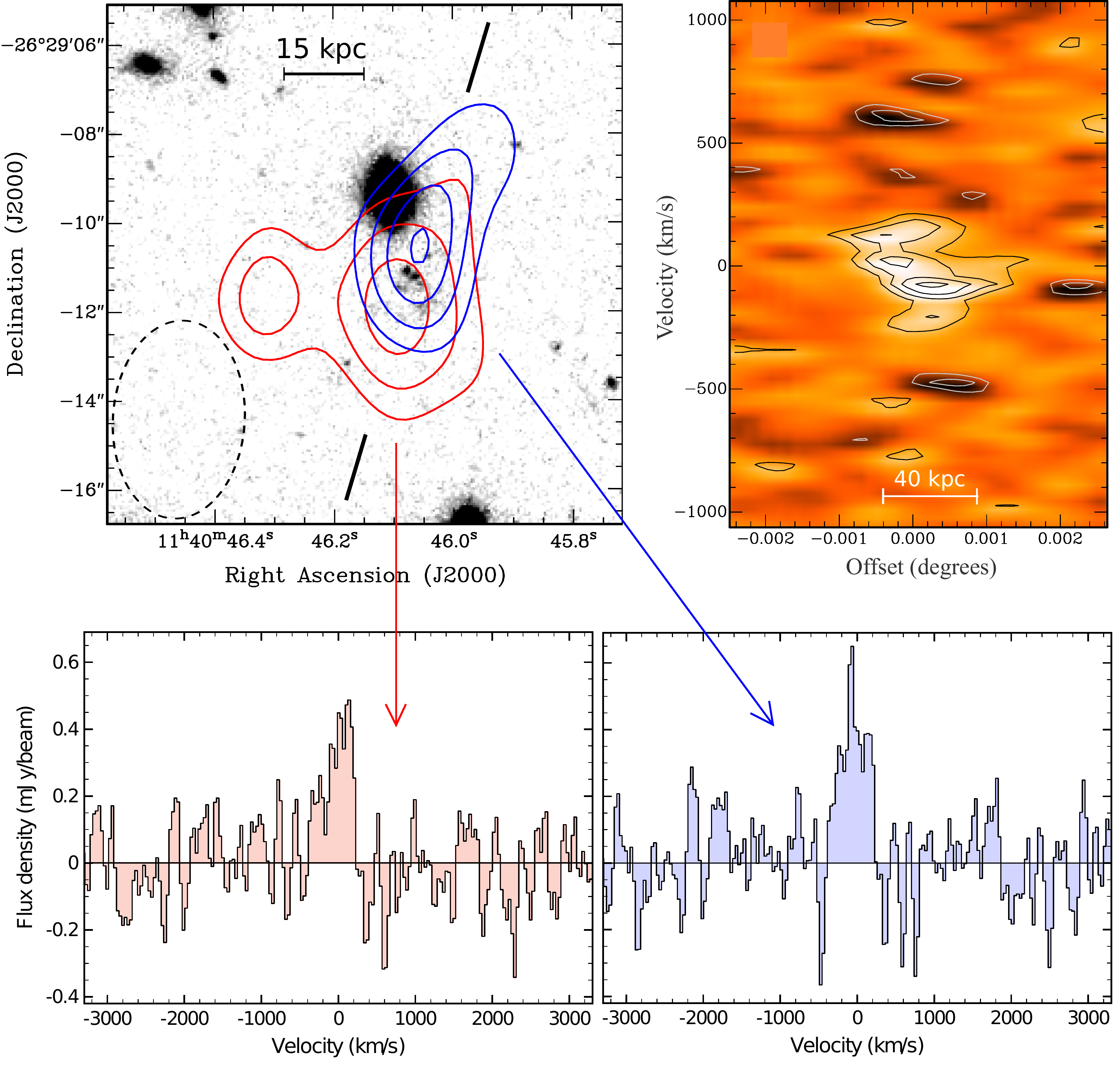}
\caption{Overview of the CO(1-0) full-resolution ATCA data. \textit{Top
left:} Total intensity image of the CO(1-0) emission across the
velocity ranges -200\,$<$\,v\,$<0$ km\,s$^{-1}$ (blue contours) and
0\,$<$\,v\,$<+200$ km\,s$^{-1}$ (red contours). Contours levels are 3,
4, 5, 6 $\sigma$ and $\sigma$\,=\,0.015 Jy\,beam$^{-1}$ km s$^{-1}$.
Following \citet{pap08}, the relative accuracy in the position of
the CO peak emission is $\lsim0.4^{\prime\prime}$. The dashed circle
indicates the half-power size of the synthesized beam. \textit{Top right:}
Position-velocity plot of the CO(1-0) emission along the line indicated
in the top-left plot (along the NE-SW direction). Contours level shown
are -4, -3, -2 (grey), 2, 3, 4, 5 (black) $\sigma$.  \textit{Bottom:}
CO(1-0) spectra of the peak emission in the blue- and red-shifted
velocity components as indicated in the plot at the top left. Note that
the spectra are not mutually independent due to the relatively large
beam and due to having velocities that over-lap.}
\label{fig:cocomponents}
\end{centering}
\end{figure*}

The most important new finding in these deeper data is that
the emission in HAE229 appears to be very extended and rotating
about the center of mass as indicated in our near-infrared imaging
(Fig~\ref{fig:cocomponents}).  The NE and SW side of the rotating
disk are separated $\sim$$1.8^{\prime\prime}$ or about $\sim$$15$~kpc
in projection, along PA~$\sim-22^{\circ}$. Their midpoint corresponds
to the peak surface brightness of the continuum emission in the
near-infrared (within the uncertainties in the astrometry of each data
set; Fig~\ref{fig:cocomponents}). The position-velocity diagram along the
blue and the red velocity components on the northeastern and southwestern
sides of the disk (Fig~\ref{fig:cocomponents}) shows a velocity gradient
of $\sim$$200$~km s$^{-1}$ in projection. The most likely explanation for
the gas properties is rotation of a gaseous disk around the centre of the
stellar mass of the system. The total CO emission is spread over a region
of $\approx$40 kpc in diameter and much of the gas is well outside the
lowest surface brightness stellar continuum emission detected in the
optical and near-infrared imaging.

\begin{figure}[ht]
\centering
\includegraphics[width=9cm,angle=0]{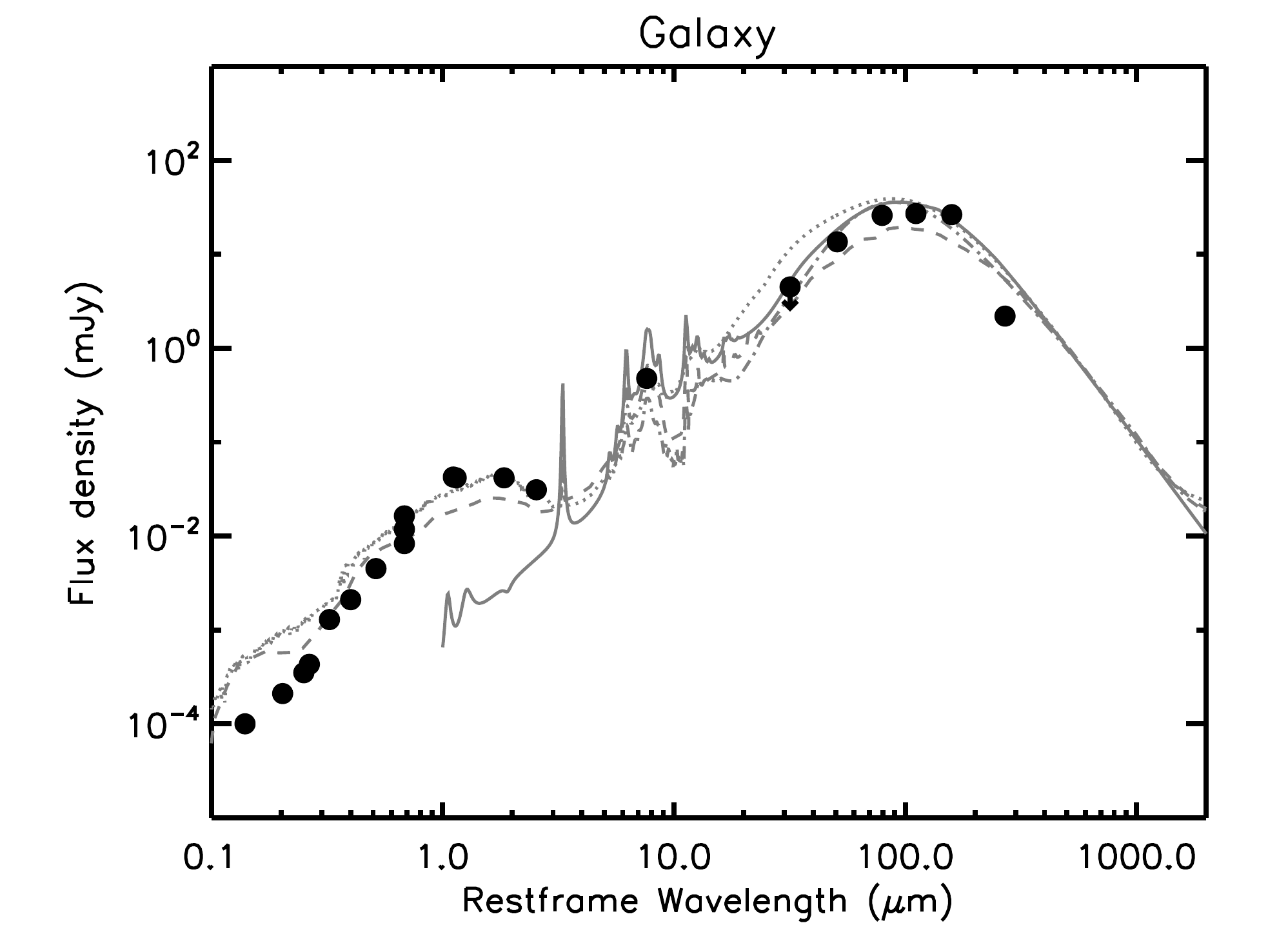}
\caption{The spectral energy distribution of HAE229 (black filled
circles). We show the template SED (solid line) of a main-sequence
galaxy at $z=2$ \citep{mag12} which is based on models \citep{dra07}. In
addition, template SEDs of observed SMGs are shown for comparison:
ALESS SMG composite \citep[dotted line][]{swi14}, average SED of 73
spectroscopically identified SMGs \citep[dashed line][]{mic10}, and an
average SED of SMGs \citep[dashed-dotted line][]{pop08}. The far-infrared
SED is matched by the different SMG composites whereas the rest-frame
optical SED shows that HAE229 is redder than the average of various
types of dusty vigorously star-forming galaxies at high redshift.}
\label{fig:fullsed}
\end{figure}

In Fig.~\ref{fig:fullsed}, we show the rest-frame UV through the FIR/submm
SED of HAE229 (Table~\ref{tab:fluxes}).  We compared the SED of HAE229
with those from a variety of other sources \citep[see][]{pop08, wei09,
mic10, mag12, hod13b, kar13, swi14}.  The far-infrared SED of HAE229 is
similar to the templates which we have choosen for making our comparison.
The template of the main-sequence galaxies resembles most closely the
far-infrared SED of HAE229.  Just like another protocluster galaxy,
DRG55 \citep{cha15}, HAE229 is significantly fainter at rest-frame
wavelengths $\la$1~$\mu$m, demonstrating that it is extremely red and is
highly dust-enshrouded.  We note that redshifted, $z=2.2$ [CII]158~$\mu$m
emission contributes to the SPIRE 500~$\mu$m flux \citep[see also][
for a more detailed discussion of this effect]{sma11}.  The uncertainty
in the SCUBA submm flux measurements of this source are evident in
the offset of the different templates relative to these measurements
\citep{ste03}. The SED of HAE229 appears to be typical of the general
population of dusty star-forming galaxies in the infrared and submm.
By integrating the infrared emission from $40-1000$~$\mu$m, we estimate
a star-formation rate of SFR$_\mathrm{IR}=555$~M$_{\sun}$~yr$^{-1}$
for HAE229\footnote{We note that if we used a Chabrier stellar initial
mass function \citep{cha03} instead of a Salpeter IMF, our estimated
SFR$_\mathrm{IR}$ would be a factor of 1.8 smaller.}.

\section{Discussion}

In extremely deep observations with the ATCA, we have found a very
extended, massive, rotating disk of cold gas in a galaxy embedded
in the protocluster surrounding the radio galaxy, MRC\,1138$-$262.  HAE229 has a high stellar mass, few 10$^{11}$
M$_{\sun}$, already in place at z=2.16. Its high CO(1-0) luminosity
suggests that it has a similar mass in cold molecular gas. HAE229 is moving
at high speed relative to the radio galaxy, MRC\,1138$-$262, over
$-$1200 km s$^{-1}$ and has a small projected separation, $\sim$250 kpc
\citep{emo13}.  \citet{dan14} showed that the far-IR and sub-mm emission
detected by \textit{Herschel}/SPIRE 250 $\mu$m and APEX LABOCA at the position
of MRC\,1138$-$262 is extended in the direction of HAE229. Although it
has a relatively high velocity relative to the radio galaxy, it has a
similar velocity to other protocluster galaxies that lie to the west of
the radio galaxy \citep{kui11}, so it is likely that HAE229 is a member
of the protocluster surrounding MRC\,1138$-$262.

HAE229 is the first CO(1-0) emitting HAE that is classified as a disk
galaxy in a (proto)cluster.  This is only the fifth detection of CO
in an HAE residing in an overdensity at high redshift. For all, with
one exception, DRG55 \citep{cha15}, the low-order CO transitions are
detected, enabling robust estimates of their total molecular gas mass to
be made without significant uncertainties due to the unknown excitation
of the gas. Three galaxies with possible low-order CO detections,
bHAE-191, rHAE-193, rHAE-213 (which is only tentatively detected),
reside in the protocluster USS~1558$-$003 at $z=2.53$ \citep{tad14}
and two of them\footnote{The third source, rHAE-213, has no reliable
FIR-measurements.}, bHAE-191 and rHAE-193, are classified as mergers.

Given the uniqueness of HAE229 -- large content of molecular gas and very
extended molecular emission -- there are several interesting questions
to address. What is the nature of this galaxy? Given that its stellar
mass is already very high, the co-moving space density of star
forming galaxies this massive is extremely low \citep[$\la$10$^{-5.5}$
Mpc$^{-3}$;][]{ilbert13} and the galaxy is also gas-rich, what
processes might prevent it from further increasing its stellar mass
substantially? Does its location in a protocluster at high redshift
affect any of the properties of its cold gas or the nature of its star
formation?  Do galaxies in protoclusters form their stars as efficiently
as field galaxies?  Are their gas fractions significantly lower as they
are in nearby cluster galaxies? We now address these questions.

\subsection{The Nature of HAE229}

A key aspect of understanding the nature of distant galaxies is
determining their mode of star formation \citep[][]{dad10b, gen10, nar15}.
Distant galaxies are hypothesized to have two modes of star formation,
a ``quiescent mode'' and a ``starburst mode''.  The quiescent mode
refers to galaxies having relatively long gas depletion time scales
(t$_\mathrm{dep} \equiv$ M$_\mathrm{H_2}$/SFR, where M$_\mathrm{H_2}$
is the mass of molecular gas and SFR is the star-formation rate)
of $0.5-2$ Gyr \citep[e.g.,][]{ler08,big08,dad10b,tac13}. This mode
has been hypothesized to be due to the long dynamical times of disk
galaxies.  Galaxies in ``starburst'' mode have short gas depletion times,
$\la$100 Myrs \citep[e.g.,][]{gre05,ivi11}, and complex morphologies and
dynamics. This mode is hypothesized to be triggered by major mergers of
two or more gas-rich disk galaxies.

In distant, $z\sim2$, protoclusters, cold molecular gas
reservoirs have only been detected in starbursts \citep{hod12, hod13a, ivi13,
rie10, tad14, wal12}. \citet{cha15} report the blind detection of star
forming molecular gas via the CO(3-2) transition of the extremely
red, main sequence galaxy DRG55 in the protocluster
HS1700$+$64. The equivalent width of the H$\alpha$ emission of this source is high suggesting that it has formed a substantial fraction of its stellar mass relatively recently
\citep{cha15}. Another four UV-selected main-sequence SFGs in
HS1700$+$64 are detected in CO(3-2) \citep{tac13}. Detecting low excitation
CO lines of normal, star-forming galaxies in overdensities is rare. This
is predominately due to the lack of systematic, targeted low-J CO studies of
galaxies in overdensities. Very recently \citet{wan16} reported the detection of 11 galaxies in CO(1-0) in a cluster at $z=2.51$. HAE229 has an exquisite multi-wavelength data
set available which enables us to investigate this galaxy in detail (see
Table~\ref{tab:fluxes} and Fig.~\ref{fig:fullsed}). From these data, there
are extensive estimates of the properties of HAE229 such as luminosities,
masses and (specific) star-formation rates (Table~\ref{tab:properties}).

\subsubsection{The molecular gas in HAE229}

How gas rich is HAE229 actually?  This is difficult to estimate accurately
due to the significant uncertainties in determining the CO
luminosity to molecular gas-mass conversion factor.  The conversion factor
depends on the gas-phase metallicity and complex gas physics \citep[see
for a review][]{bol13}.  Ignoring the complexities of determining the
appropriate conversion factor, we focus on using its metallicity
dependence to constrain the conversion factor for HAE229.  \citet{kur04b}
find a log [NII]/H$\alpha$ ratio of $-$0.47, typical of star-forming
galaxies at these redshifts \citep[there is no evidence for a strong
AGN contribution;][]{ogl12, koy13}. The [NII]/H$\alpha$ ratio implies a
metallicity, 12+log(O/H)~$\approx8.6-8.8$ \citep{kur04b, pet04, den02, man10,
mai14}.  We apply the gas-to-dust ratio method, ``$\delta\mathrm{GDR}$'',
for estimating the CO-to-H$_{2}$ gas conversion factor, $\alpha_\mathrm{CO}$ \citep{ler11,mag11,mag12}.  This method relies on an estimate of
the dust mass for which we find M$_\mathrm{dust}=3.5\times10^{9}$~M$_{\sun}$ \citep[using the method of][]{cas12}. Using the equation
log~$\delta\mathrm{GDR}=(10.54\pm1.0)-(0.99\pm0.12)\times(12+\mathrm{log(O/H)})$
\citep{mag12}, we estimate the CO-to-H$_{2}$ conversion factor, obtaining, $\alpha_\mathrm{CO}$
= $4.7-6.9$~M$_{\sun}$~pc$^{-2}$~(K~km~s$^{-1}$)$^{-1}$. The
range of $\alpha_\mathrm{CO}$ we give here depends on the method
\citep{den02,pet04,man10}. Using a metallicity-dependent conversion
factor suggests that HAE229 is similar to local and distant normal
main-sequence disk galaxies.

Given all of the uncertainties in estimating the conversion factor,
it is not entirely clear that if basing an estimate directly on
the dust mass and metallicity is completely appropriate or robust.
Another way of estimating the conversion factor is to use an independent
estimate of the gas mass and simply scale the CO luminosity to give
the same gas mass. For example, we can use the empirical calibration
of the long wavelength dust continuum emission to estimate the total
gas mass \citep{sco16}.  Using the $850~\mu$m flux density,
we find a total gas mass, M$_\mathrm{gas}\sim$ 3 $\times$ 10$^{11}$
M$_{\sun}$. This estimate includes an unconstrained contribution from
HI, so this provides only an upper limit to the conversion factor and
implies M$_\mathrm{mol}$/L$^{'}_\mathrm{CO(1-0)}\la6$.

All of our estimates favor a relatively high CO conversion factor.  HAE229
has a high infrared luminosity, it is a SMG (although the detection at
850~$\mu$m is not highly significant).  Many authors favor a low value of
the conversion factor for luminous infrared sources \citep[0.8 M$_{\sun}$
(K km s$^{-1}$ pc$^{2}$)$^{-1}$; ][]{sol05}.  With these caveats in mind,
we adopt a conversion factor of 4 \citep[typically for high-z disk-like
galaxies, see e.g.,][]{dad10a}, implying a total cold molecular gas mass,
M$_\mathrm{mol}$ = 2.0$\pm$0.2$\times$10$^{11}$ M$_{\sun}$.  Adopting a
low value generally appropriate for luminous galaxies like HAE229 would
lead to conflicts with other estimates of gas masses given previously
which do not rely on the conversion factor directly \citep[e.g.,][]{cas12,
sco16}.

\begin{figure*}[!t]
\centering
\includegraphics[width=7.3cm,angle=0]{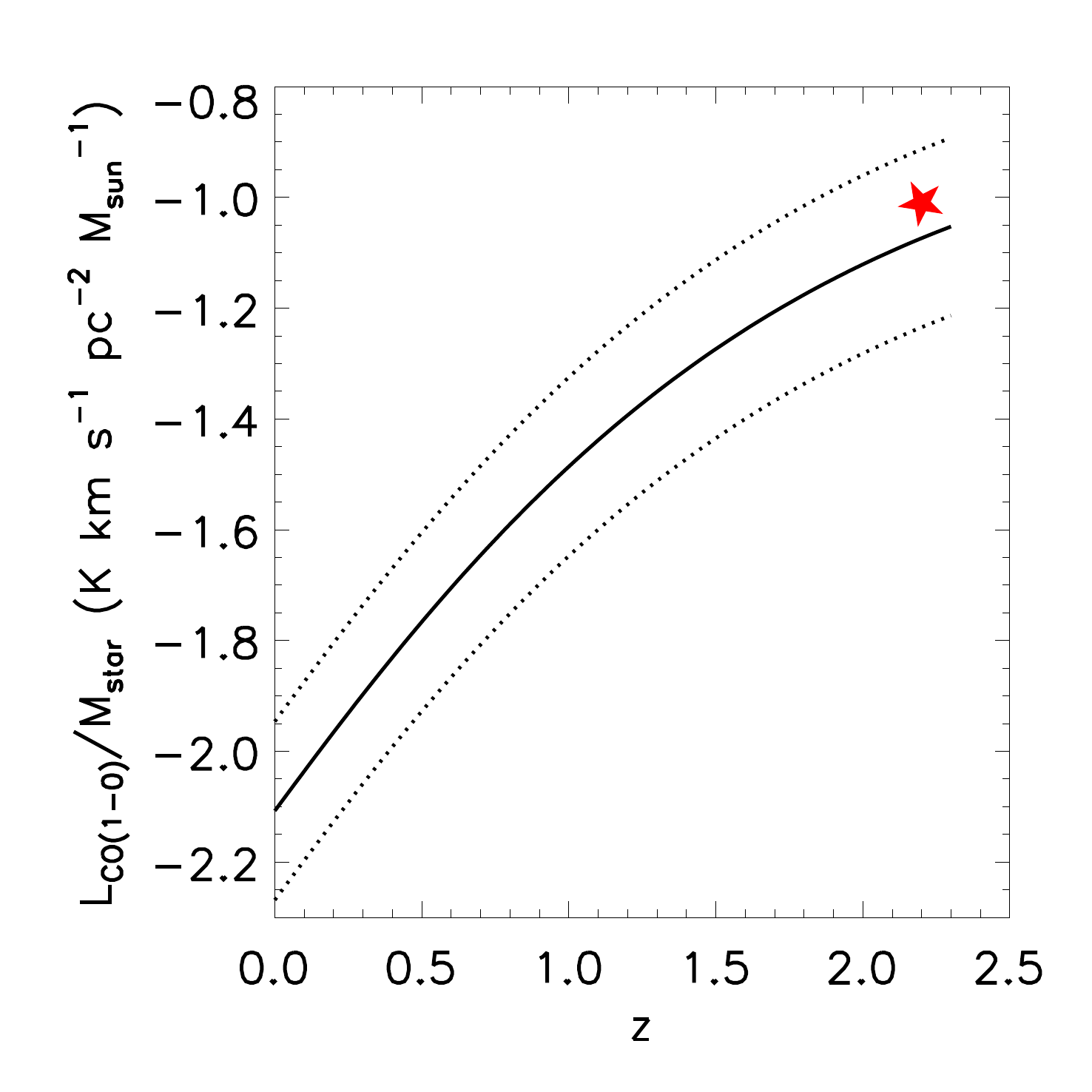}
\includegraphics[width=7.3cm,angle=0]{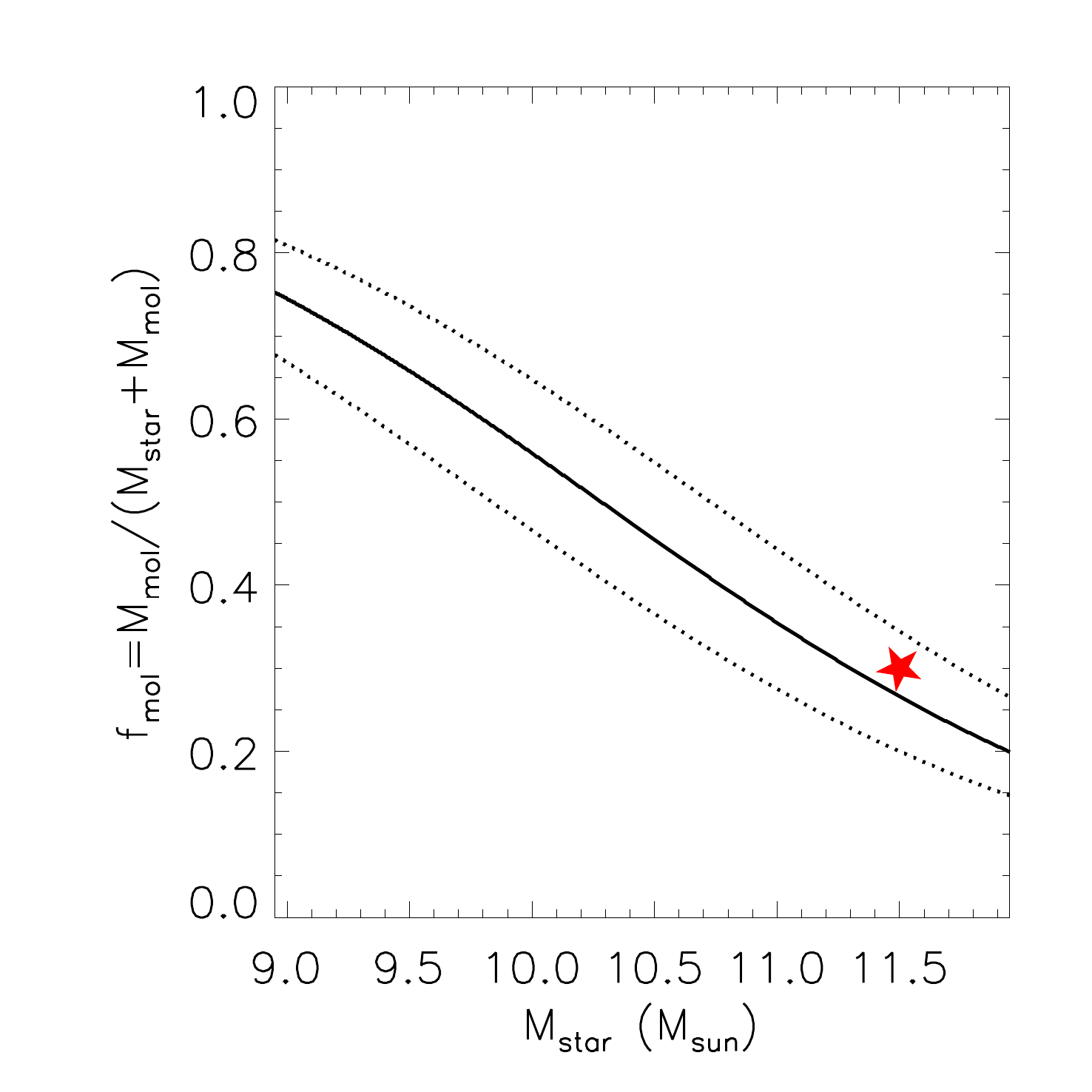}
\caption{\textit{Left panel:} Ratio of the CO emission and stellar mass
in units of K km s$^{-1}$ pc$^{-2}$~M$_{\sun}$$^{-1}$ versus redshift . The
solid and dotted line shows the median and 1$\sigma$ dispersion of this
ratio for main-sequence galaxies with a mass of $5\times10^{11}$~M$_{\sun}$.
The mass we have chosen for this comparison is approximately the
stellar mass of HAE229. \textit{Right panel:} Molecular gas fractions
and $\pm$1-$\sigma$ dispersion (solid and dotted lines respectively)
for main sequence galaxies at $z=2.2$. The lines shown in both panels are from
models discussed in \citet{sar14} to explain the evolution of the main
sequence of galaxies. The red star represents the estimated position of
HAE229 in both panels.}
\label{fig:gasproperties}
\end{figure*}

Using the adopted conversion factor and CO luminosity,
we estimate a molecular gas fraction, f$_\mathrm{mol}$ =
M$_\mathrm{mol}$/(M$_\mathrm{mol}$+M$_{\star}$), of $\sim$30\%
for HAE229.  In Fig.~\ref{fig:gasproperties}, we show the expected
redshift evolution of the ratio between CO luminosity and stellar mass
based on the empirical scaling relations for a typical main-sequence
galaxy \citep[see][]{sar14}. HAE229 lies well on the predicted relation for main sequence galaxies at $z=2.2$. The same applies
when comparing HAE229 to the expected average variation of the gas
fraction with stellar mass for galaxies at the redshift of HAE229
(Fig.~\ref{fig:gasproperties}).

\subsubsection{Timescales: Do the disk dynamics limit the star formation
in HAE229?}

A quantitative way to obtain insights into the ``mode of star
formation'' in galaxies is to derive the specific star-formation
rate and ``star-formation efficiency''. If HAE229, for its stellar
mass and redshift, is a typical main-sequence galaxy, we would expect
its star-formation rate, SFR$_\mathrm{MS}$ $\approx$ 460 M$_{\sun}$
yr$^{-1}$ \citep{tac13}.  We derived a far-infrared based star-formation
rate of SFR$_\mathrm{IR}=550~$M$_{\sun}$~yr$^{-1}$ (\S~\ref{sec:results})
which falls within 20\% of SFR$_\mathrm{MS}$ at its redshift, i.e. well
within the 1-sigma scatter of the relation between SFR and stellar mass
\citep[e.g.,][]{mag12}. HAE229 is a distant, massive main-sequence galaxy.

Our estimate of the star-formation rate (\S~\ref{sec:results}) and
molecular gas mass implies a gas depletion time, t$_\mathrm{dep}$
$\approx360$ Myrs \footnote{We note that, if we adopt a lower
CO-to-molecular gas mass conversion factor, say $\alpha_{CO}=1$, the gas
depletion time for HAE229 would be about 100 Myr, see \citep{cas16}.}. The
range of star-formation rates that have been estimated for HAE229
implies a systematic uncertainty in the gas depletion time of about
a factor of 2.  Our estimate for HAE229 is similar to the depletion
times derived for star-forming galaxies at $z=1-3$ that are not
undergoing a major merger \citep{dad07,gen10,rod11}, but are up to an
order-of-magnitude higher than the typical depletion times derived for
SMGs \citep{gre05,tac08,ivi11,rie11}, although see also \citet{swi06}
for exceptions. This is another indication that HAE229 is a normal high-z
star-forming galaxy.

The dynamical time (rotational), t$_\mathrm{dyn}$~=~2$\pi$r/(v/sin
i)$~=~610$ Myrs $\times$~r$_\mathrm{10 kpc}$/(v/sin i)$_\mathrm{100 km
s^{-1}}$, where r$_\mathrm{10 kpc}$ is the radius in units of 10 kpc,
(v/sin i)$_\mathrm{100 km s^{-1}}$ is the inclination corrected velocity
in units of 100 km s$^{-1}$, and i, is the inclination.  In order to
constrain the inclination, we estimate the rotation speed needed for
the disk to be centrifugally supported, i~$\la$~30$^{\circ}$, and (v/sin
i)~$\approx$~500 km s$^{-1}$. At 10 kpc, the dynamical time is $\sim$120
Myrs. But the gas may be extended over a region up to twice this large and
thus, in the case of HAE229, t$_\mathrm{dep}$/t$_\mathrm{dyn}\approx2-3$.
Finding that the gas depletion time scale is a factor of a few greater
than the dynamical time is interesting in light of the gas distribution
relative to that of the on-going star formation in HAE229. The molecular
gas appears to spread over a larger area than the stellar continuum. Thus
it is clear that we must consider the role the high angular momentum
of the extended disk plays in regulating future star formation. While
we do not have the resolution to investigate the detailed kinematics
of the disk, the fact that the gas distribution is so extended and
has significant angular momentum makes us wonder if it could be that
HAE229 will not be able to support its on-going star formation for a
gas-depletion time scale but will use up the gas in the central regions
most quickly.  In this case, what limits the time over which intense star
formation takes place, is not when the gas over the disk is exhausted,
but how long it takes for the gas to dissipate its angular momentum and
to move from the outer to the inner disk. Perhaps it is the high angular
momentum of the gas and the rate at which the extended gas in galaxies
dissipates angular momentum that ultimately regulates the duration of
intense star formation \citep[see][]{leh15}.

In order to sustain the SFR of HAE229 beyond a dynamical time,
$\approx$500 M$_{\sun}$ yr$^{-1}$ of gas must be accreted into its
central region where the on-going intense star formation is
concentrated. While, there is a large reservoir of gas in the
outskirts, to feed the regions of growth in the inner disk, it must
dissipate a significant amount of angular momentum, about a factor of
2 or more within a dynamical time.  It is not clear how it would
accomplish this feat, as there are no signs of a bar or an intense
interaction with a companion galaxy.  Asymmetries in the mass
distribution which would generate torques on the gas enabling the
extended gas to dissipation energy and angular momentum, are not
readily apparent \citep[e.g.,][]{gav15}. Given these circumstances,
since a detailed analysis of this is beyond the scope of the paper, we
simply raise the point that using the gas depletion time as an
indicator of the duration of the star formation or its regulation, is
at best naive, at worse, misleading. Higher resolution observations of
the molecular gas might allow us to investigate this idea in more
detail. Finally, we note that analysis of \citet{spi15} shows that for
a sample of distant galaxies with detection in the low-J CO lines, the
CO emitting effective radii are larger than the effective radii of the
star formation. Thus our claim that dynamical effects may be important
in limiting the gas supply of galaxies could be generally true for the
population of distant star-forming galaxies.

\subsection{Previous detections of the cold molecular gas in $z>0.4$
cluster galaxies} \label{sec:previous}

Motivated by the CO(1-0) detection of the protocluster member HAE229, it
is interesting to investigate the dependency of the physical properties
of the molecular gas in (proto)cluster galaxies beyond $z\sim 0.4$,
thus extending the previous study by \citet{jab13} of molecular gas
in clusters members to higher redshifts.  Therefore, we searched
the literature for previous detections of the cold molecular gas in
(proto)cluster galaxies at $z>0.4$. We restrict our search exclusively
to CO(1-0) or CO(2-1) detections as we want robust estimates of
the total amount of molecular gas masses \citep[cf.][]{cas16}. Using
high-order CO transitions, from CO(3-2) upwards, can lead to significant
underestimates of the true gas mass \citep[e.g.,][]{dan09}. High-order
transitions trace the dense, high excitation gas closely associated
with star-forming regions and feedback.  The low order transitions,
CO(1-0) and CO(2-1)\footnote{We assume a ratio of 1 between the CO(2-1)
and CO(1-0) luminosity.}, trace the lower density gas, which is a better estimate of a galaxy's potential to form
stars \citep[e.g.,][]{ivi11,emo15a}. Including HAE229, we find 24 CO(1-0)
and/or CO(2-1) bright galaxies in 12 different galaxy overdensities
beyond $z>0.4$: seven lie between $0.4<z<1.0$, three between $1.0<z<2.0$,
seven between $2.0<z<3.0$, one at $z=3.1$, three at $z=4$ and two beyond
$z=5$, see Table~\ref{clusterco} for details. Our (proto)cluster sample
is dominated by high-z sources \citep[cf.][]{jab13}. The environments
probed are both clusters and protoclusters at different evolutionary
stages. The number of detections of the cold molecular gas is still low
compared to field galaxies and none of these clusters has a large number
of detections. The infrared luminosity of this sample range between
L$_{\rm IR}$~$\sim$~$5-400$~$\times$~10$^{11}$~L$_{\sun}$. Based on
this compilation, we conclude that even high L$^{\prime}_\mathrm{CO}$
galaxies exist in high density fields \citep[cf.][]{jab13}. Below
$z=1$ all of the cluster sources have a disk-like star-formation
mode. Thus, we can extend the relation between L$_\mathrm{IR}$ and
L$^{\prime}_\mathrm{CO}$ for cluster galaxies both in L$_\mathrm{IR}$
and L$^{\prime}_\mathrm{CO}$ compared to \citet{jab13} who
were restricted up to L$_\mathrm{\rm IR}<10^{12}~L_{\sun}$ and
L$^{\prime}_\mathrm{CO}<10^{10}$~K~km~s$^{-1}$~pc$^{2}$. They concluded
that the frequency of high L$^{\prime}_\mathrm{CO}$ galaxies in clusters
is lower than in the field implying that the molecular gas content of massive
galaxies depends on environment.

With our compilation of 24 detections beyond $z>0.4$ we find that
above this redshift, as we now discuss, the cluster environment does
not influence strongly the molecular gas content and other properties
of galaxies in the early universe. We note that our sample at
$z>0.4$ is dominated by protoclusters members. Thus, there could be
a bias in our compilation as high-z protocluster are, by definition,
the early stages in the evolution of clusters \citep[see][for more details on
protoclusters]{ove16}.

\subsubsection{CO(1-0) Line width-luminosity relation}

The baryonic Tully-Fisher (T-F) relation can be understood as a natural
relationship between angular momentum and mass of centrifugally-supported
disk galaxies. In the standard cosmological model, a Tully-Fisher
relationship results naturally if the fraction of the angular momentum
and mass of the baryons are both fixed fractions of the angular momentum
and mass of the dark matter halos themselves \citep{mo98}. The classical
Tully-Fisher relationship is measured using the rotation speed of
the disk and the total stellar content or luminosity of the disk or
total baryonic mass \citep[e.g.,][]{mcg15}. In a twist in the study of
specific ``T-F-like relations'', \citet{bot13} found a relationship
between CO luminosity, L$^{\prime}_\mathrm{CO}$, and line width,
FWHM, for SMGs \citep[cf.][]{car13,sha16}. \citet{got15} confirm
this finding. \citet{bot13} interpret this as a uniform ratio of the
gas-to-stellar contribution to the dynamics of CO-bright regions. They
model this trend as a relationship between the molecular gas mass and the
velocity necessary for centrifugal support of a rotating disk. These
authors further suggest that the low scatter, lower than that expected
for randomly inclined disks, is due to the fact that galaxies at high
redshift are geometrically thicker than their low redshift counterparts
and that the molecular gas is a significant fraction of the total
baryonic mass of the galaxies.  Taking this relationship even further,
despite its large scatter, \citet{har12} and \citet{zav15} use this
relation to de-magnify \citep[cf.][]{ara16} \textit{Herschel}-selected lensed
high-z galaxies \citep[e.g.,]{neg10, vie13, mes14}.

Our discovery of a large rotating gas disk motivates us to investigate
whether a Tully-Fisher relation exists for the CO emitting gas for
a wider range of galaxies beyond just SMGs.  Here we include high-z
disk-like galaxies and search for differences between field and cluster
galaxies. A difference is to be expected because additional processes
operate in environments with higher galaxy and inter-galactic gas
densities \citep[e.g.,][]{jab13}.

\begin{figure}[th]
\centering
\includegraphics[width=8.0cm,angle=0]{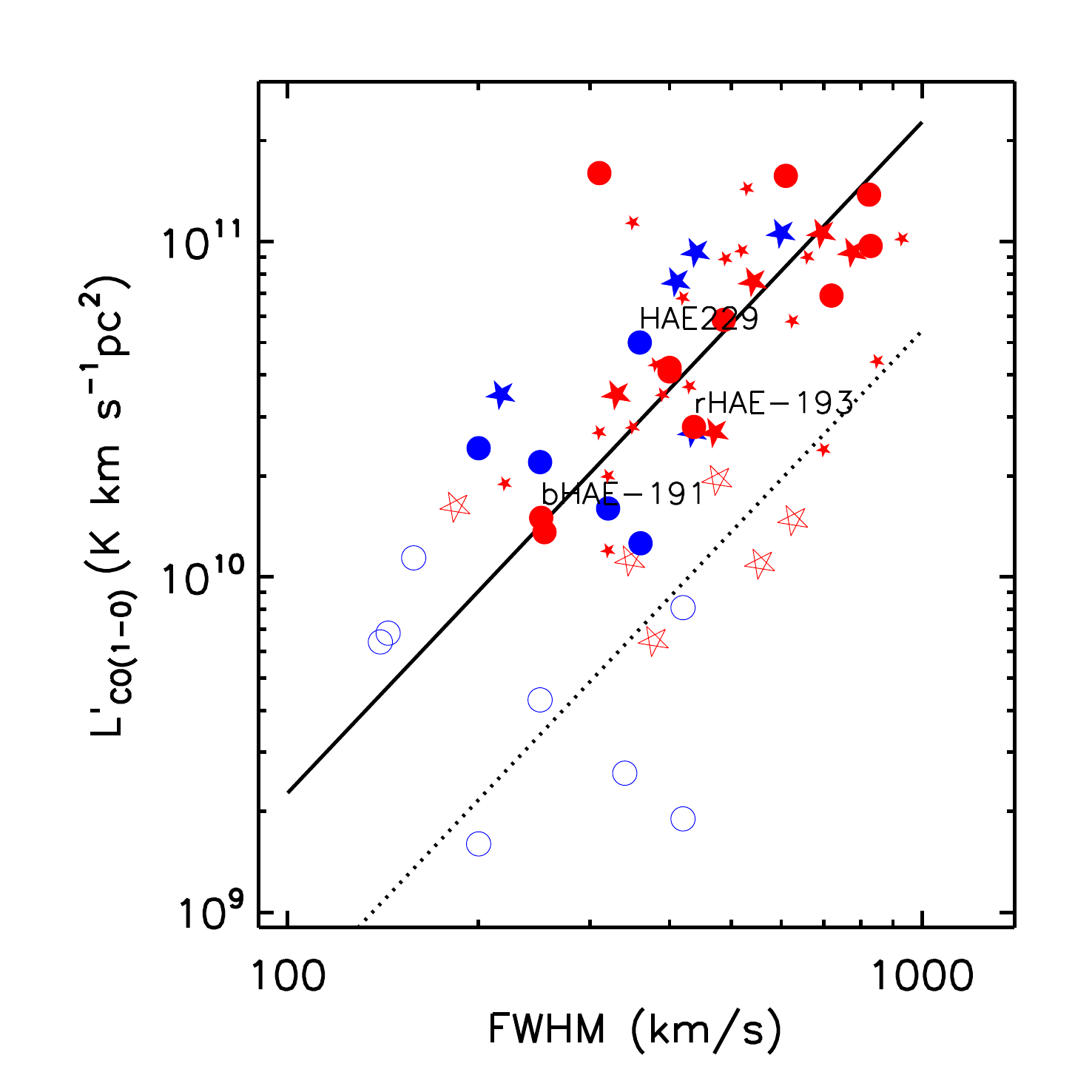}
\caption{The relation between FWHM of the CO(1-0) line and
L$^{\prime}_\mathrm{CO}$ for cluster (circles) and field galaxies
(stars). High redshift sources, those with z$>$1, are indicated by
filled symbols while sources with z$<$1 are indicated by hollow ones.
The color-coding used in the figure is as follows: intermediate-z ULIRGs
\citep[red hollow stars][]{mag14}, normal SFGs at $z=1.5$ \citep[blue
filled stars][]{dad10a,ara14}, and SMGs \citep[large and small red filled
stars from][respectively]{ivi11,bot13}. References for cluster sources
can be found in \S~\ref{sec:previous} and Table~\ref{tab:cocluster}. The
mode of star formation for each source is color-coded as, blue for
``disk-like'' and red for ``starburst''.  The solid line shows the
relation from \citet{bot13}.  The dashed line assumes the typical radius
for disk galaxies. Both high-z population, dusty starbursts and disk-like
galaxies, show a unique correlation between CO luminosity and line width
with significant scatter ($\sim$0.3 dex).}
\label{fig:tf}
\end{figure}

Our sample of (proto)cluster galaxies is a mix of disks and starbursts
culled from the literature (see Table~\ref{clusterco}).  We find
that most of the cluster sources beyond $z=0.4$ follow the same
L$^{\prime}_\mathrm{CO}$-FWHM relation as found for SMGs over the same
redshift range \citep{bot13, got15} but some of the galaxies in our sample
lie below this relation \citep[Fig.~\ref{fig:tf}; see also][]{car13}.
Interestingly, the cluster galaxies below the relation are at low-z and
overlap with the \textit{Herschel} (U)LIRG field sample at intermediate
redshifts \citep{mag14}.  In addition, we show unlensed, high-z field
galaxies detected in CO(1-0) \citep{ivi11,bot13} and normal star-forming
galaxies selected via 3 color imaging \citep[B-, z-, and K-b imaging
-- BzK galaxies][]{dad04} that are detected in CO(1-0) and/or CO(2-1)
\citep{dad08,dad10a,ara10,ara14}.  Overall, we find that both normal
SFGs and dusty starbursts at high-z follow the same relation as SMGs
\citep{bot13}, while low and intermediate redshift ULIRGs perhaps have
a weak trend offset from the high-z trend. 

It is interesting that the SMGs and the disk galaxies in the field and
protocluster environments appear to follow the same approximate relation.
The explanation in \citet{bot13} of L$^{\prime}_\mathrm{CO}$--FWHM
trend for distant galaxies is based on the argument that the galaxies in
this relation are centrifugally-supported disks and the molecular gas
constitutes a large fraction of the total mass. However, \citet{bot13}
made two assumptions that may not be appropriate for all the galaxies
on this relation. They assumed a conversion factor of 1 M$_{\sun}$ (K
km s$^{-1}$ pc$^{2}$)$^{-1}$ and a radius of 7 kpc to make the relation
match the data.  If the disks have a radius of a factor of 1.9 times
higher than starbursts and a higher $\alpha_\mathrm{CO}$ then the two
relations, the ones for high and low redshift,  would be the same. SMGs
and normal SFGs may have the same parent population \citep[e.g.,][]{hay11,
hay12,bet15}; a finding that can perhaps be explained by a variation
in the gas fraction: the gas fraction decreases with the total galaxy
mass at a given redshift and decreases with decreasing redshift at a
given total mass \citep[e.g.,][]{sar14}. Since the baryons dominate the
potential within the stellar mass distribution of halos, for a galaxy
at a given baryonic mass (meaning constant FWHM if the gas motions are
virialized), the gas fraction will decrease for galaxies with higher
stellar mass \citep[e.g.,][]{sar14}. Thus, we may expect zero-point of
the L$^{\prime}_\mathrm{CO}$--FWHM to decrease with decreasing redshift
(if such a relation exists at all at lower redshifts).

The most interesting finding in this analysis, however, is not the
significance of the trend, but that we do not find any difference between
cluster and field galaxies in the FWHM-L$^{\prime}_\mathrm{CO}$ plane
for high-redshift sources.  The trend is independent of the environment
for the galaxies in these samples.  This implies that for both field and
(proto)cluster galaxies their gas content (gas as a total fraction of
their total baryonic content) and dynamics are, within a wide scatter,
similar.  At lower redshifts, \citet{jab13} found that the field and
cluster galaxies have different gas content for the equal stellar mass.
This suggests that what ever processes dominate the regulation of the
gas content of galaxies in low redshift clusters do not operate as
efficiently at redshifts $\ga$0.5-1.

However, intermediate-z sources do not follow this relation.  Why?
ULIRGs, the powerful infrared emitters at low and intermediate redshifts,
are dominated by systems with strong random, non-virialized motions.
Some SMGs, on the other hand, appear to have dynamics that are ordered,
perhaps virialized \citep[e.g., GN20; ][]{hod12,hod13a} whether or not
they lie in clustered environments or not.  This agrees with the fact
that SMGs are not simply scaled-up versions of local and intermediate
redshift ULIRGs \citep[e.g.,][]{swi14}.

\begin{figure}[th]
\centering
\includegraphics[width=8.0cm,angle=0]{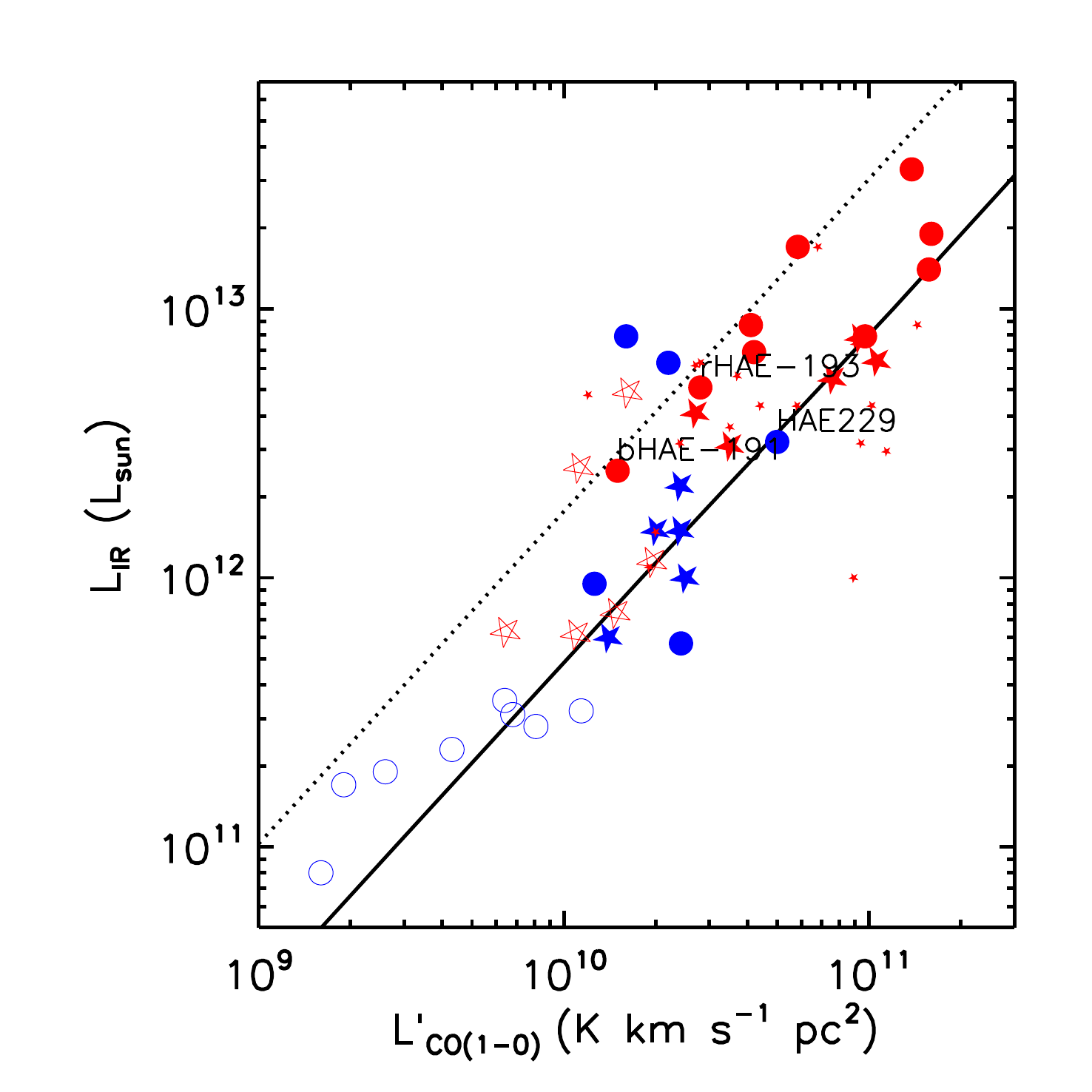}
\caption{The relationship between L$_{\rm IR}$ and
L$^{\prime}_\mathrm{CO}$ -- the integrated Schmidt-Kennicutt law --
from our compilation of CO-bright bright galaxies lying in overdensities
or protoclusters (circles, same encoding as in Fig.~\ref{fig:tf};
Tab.~\ref{tab:cocluster}). The three CO-bright HAEs are indicated
individually. HAE229 is clearly separated from the two other HAEs with
CO detections \citep{tad14}, potentially indicated that red and blue
HAEs have a wide range of CO luminosities. Small numbers of HAE with
CO detections prohibit us from deciding if this is a real astrophysical
effect. Solid and dashed lines are the best-fitting relations for normal
star-forming and starburst galaxies \citep{sar14}. At least within this
small sample, no clear bi-modality in the star-formation efficiency is
	observed \citep[cf.][]{dad10b, gen10}.}
\label{fig:kslaw}
\end{figure}

\subsubsection{Integrated Schmidt-Kennicutt relation}

To further investigate whether or not being located in a galaxy
overdensity impacts the gas content and star-formation rate of
galaxies, we now study the relationship between star-formation
rate and gas content -- the integrated Schmidt-Kennicutt relation.
The ratio of the SFR and M$_\mathrm{mol}$ can be thought of as a
star-formation efficiency, that is, the conversion efficiency of
gas into stars.  We can cast this relationship directly between two
observational quantities -- one which is proportional to the SFR,
the infrared luminosity, L$_\mathrm{IR}$, and the other, to the mass
of molecular gas, the CO luminosity, L$^{\prime}_\mathrm{CO(1-0)}$. If
the conversion between L$_\mathrm{IR}$ and the SFR and the conversion
between L$^{\prime}_\mathrm{CO(1-0)}$ and the gas mass does not depend
on the characteristics of the galaxies themselves, then this ratio,
L$_\mathrm{IR}$/L$^{\prime}_\mathrm{CO(1-0)}$ will be proportional to
the ``star-formation efficiency''.  Any differences found, would then
be attributable to the differences in either the conversion between
CO luminosity and the gas mass, $\alpha_\mathrm{CO(1-0)}$, which would
indicate that the excitation of the cold molecular gas depends on the
characteristics of the galaxies themselves.  Alternatively, if there are
not such dependencies in $\alpha_\mathrm{CO(1-0)}$, any differences may
be attributable to the rate at which gas is converted into stars, i.e.,
the ``star-formation efficiency''.

Several studies have found a bi-modality in the integrated
Schmidt-Kennicutt relation \citep[e.g.,][]{dad10b,gen10}.
This bi-modality has been attributed to starburst galaxies having
a higher star-formation efficiency than the normal galaxies.
This is despite having conversion factors that would suggest
that starburst galaxies are less gas-rich in proportion to their
CO luminosities.  In Fig.~\ref{fig:kslaw}, we investigate the
L$_\mathrm{IR}$/L$^{\prime}_\mathrm{CO}$ ratio for our compiled
CO-bright cluster sample, intermediate-z ULIRGs \citep{mag14}, normal
SFGs at $z=1.5$ \citep{dad10a} and SMGs \citep{ivi11,bot13}\footnote{We
note that in absence of CO(1-0) observations, \citet{bot13} converted
measurements of transitions higher or equal than CO(2-1) into the CO(1-0)
transition.}. We find that HAE229 lies on the relation of disk-like
galaxies whereas the other CO(1-0) detected HAEs \citep{tad14} lie on
the starburst relation \citep{sar14}.  At infrared luminosities below
L$_\mathrm{IR}$~=~2~$\times$~10$^{12}$~L$_{\sun}$, almost all galaxies
follow the disk-like star-formation relation. Above this luminosity both
relations, ``disk-like'' and ``starburst'', are populated.  Interestingly,
as previously pointed out, SMGs do not follow only the relation
for ULIRGs and dusty starbursts but also lie at the region expected
for disk-like galaxies \citep[e.g.,][]{swi11, hod12}, demonstrating
that this source population is not homogeneous \citep{hay11, hay12}.
Our analysis suggests that the simple dichotomy in the galaxy population
based only on the star-formation efficiency is perhaps more complex
than previous interpretations \citep[e.g.,][]{dad10b,gen10}.  We do not
find a difference of the L$_\mathrm{IR}$ and L$^{\prime}_\mathrm{CO}$
relation between cluster and field galaxies. We therefore conclude, with
the present data quality, that lying in a denser environment at high
redshifts does not significantly alter the star-formation efficiency or
the molecular gas excitation conditions in galaxies.

\subsubsection{Are the environmental drivers of galaxy evolution efficient
in the early Universe?}

We have found that for gas-rich galaxies with redshifts larger than about
0.4, there does not appear to be an environmental dependence for the gas
content, the star-formation efficiency, or on excitation conditions of the
diffuse molecular gas as probed by low-order CO line transitions.  This
finding could be a sign that typical physical processes in local clusters
that are responsible for depleting the content and altering the physical
conditions of the gas in galaxies, like harassment, tidal stripping,
and ram-pressure stripping \citep{moo96, vol01b, gne03a, gne03b,bos14a}
do not operate efficiently at high-redshift \citep{jab13,hus16}. At
higher redshift, for a given cluster mass, the internal galaxy-galaxy
velocity dispersions of a protocluster will be higher than for local
clusters. The galaxies surrounding the radio galaxy have extremely high
relative projected velocities, of-order 1000 km s$^{-1}$, suggesting
that the protocluster is already massive \citep{kui11}.  The processes
that are likely to affect both the gas content and evolution of galaxies
in clusters have very different dependences on the relative velocity
of a galaxy as it moves through the population of cluster galaxies
and cluster potential \citep[e.g.,][]{moo96, vol01b, gne03a, gne03b}.
Both the harassment and tidal stripping rates are significantly lower
if the galaxy, \textit{(1)} has no close companions with small relative
velocity; \textit{(2)} has a high velocity relative to the mass center
of the cluster; or \textit{(3)} is not moving through regions of the
cluster with a high galaxy volume density. In the case of HAE229, it
has a high projected velocity relative to the radio galaxy, which if
we assume it represents the center of mass of the cluster, implies that
strong tidal stripping is very unlikely.  It is also at a relatively large
projected separation, about 250~kpc.  While small compared to the likely
virial radius of a massive dark matter halo at $z\sim2$ \citep[see ][
and references therein]{kui11}, it means that HAE229 is not feeling the
influence of the total mass of the cluster. Also at high velocities,
the timescale for harassment by other galaxies is very short and thus
unlikely to be very effective in removing gas. These effects will be
further reduced if protoclusters galaxies have not yet virialized \citep {kui11} and
at least in the case of HAE229, has yet to make its closest approach to
the center of the protocluster potential.

However, the high velocity of HAE229 relative to the radio galaxy
\citep[$\sim$1200 km s$^{-1}$;][]{emo13}, should, in principle,
lead to very efficient ram pressure stripping.  Ram pressure
stripping is proportional to average gas volume density,
$\left<\rho_\mathrm{w}\right>_\mathrm{v}$ multiplied by the
square of relative velocity of the galaxy in the potential,
$v_\mathrm{rel}^{2}$.  The volume-averaged gas density,
$\left<\rho_\mathrm{gas}\right>_\mathrm{v}$ = $\phi_\mathrm{gas, v}
\rho_\mathrm{gas}$, where $\phi_\mathrm{gas, v}$ is the gas volume-filling
factor and $\rho_\mathrm{gas}$ is the density of the gas responsible for
stripping the galaxy. The key to ram pressure operating effectively is
for the gas within the cluster potential to have both a relatively high
density and high filling factor \citep[e.g.,][models assume a high volume
filling factor, $\phi_\mathrm{gas, v}$=1 for hot halo gas and thus neglect
this term]{gne03b}. The ram pressure must overcome the restoring force
provided by the gravitation potential at the disks surface.  The higher
velocities of galaxies in high redshift (proto)clusters compared to low
redshift clusters \footnote{The cluster velocity dispersion, $\sigma$,
scales as the virial velocity, (G M$_v$/r$_v$)$^{1/2}$, where M$_v$
and r$_v$ are the virial mass and radius, respectively. This relation
scales as ($\Delta$(z)/2)$^{1/6}$ (G H(z) M$_v$)$^{1/3}$, where the
mean density of the halo is $\Delta$(z) halo times the critical density,
H(z) is the Hubble constant, and G is the gravitational constant.  $\Delta$(z) 
increases slowly with increasing z, the Hubble constant H(z) increases
rapidly with z, and thus so does the velocity of galaxies in halos of
constant virial mass as a function of redshift.}, for a constant mass,
suggests that the impact of ram pressure stripping should be enhanced,
not diminished in distant clusters. Gas at high temperatures dominate the
inter-cluster medium (ICM) and the mass budget of low redshift clusters
\citep[e.g.,][]{lag13}. The hot ICM gas has densities of $\sim$10$^{-3}$
to 10$^{-1}$ cm$^{-3}$, temperatures of $\sim$10$^{7}$ to 10$^{8}$ K,
and a near unity filling factor \citep[e.g.,][]{san09}. The densities,
unity filling factor, and high velocity dispersion of the galaxies
is why ram pressure stripping is so effective in low redshift clusters
\citep[e.g.,][]{vol01a,vol01b,fum14,bos14b,bos16}. The near unity filling
factor of the hot gas means that as the galaxy is moving through the ICM,
it feels a steady wind of material with slowly varying density depending
on where it is relatively to the cluster center. Since we do not find a
strong impact on the gas content like that seen in local cluster galaxies,
it must be that the dominant gas phase in distant protoclusters either
has a density which is much less but still has a filling factor of unity
or that it has a much lower volume filling-factor than in local clusters.

The ram pressure due to the ICM is counterbalanced by the restorative
force of the disks gravitational potential. The restorative force is
proportion to the radial distribution of the combined stellar and gaseous
surface mass density. In other words, galaxies with compact and roughly
equal gaseous and stellar mass surface densities, have maximal restoring
forces and are resistant to ram pressure stripping. But galaxies at high
redshift, and HAE229 in particular, have very extended gas distributions.
The disk of HAE229 extends well beyond its stellar light. We would
expect its outer gas disk to be easily stripped.  For that matter,
even tidal stripping should be effective in such extended disks.
Why has HAE229 not been stripped? Again, this would be consistent with
the mass content of the IGM of high redshift clusters not having a
high volume filling-factor gas of sufficient volume-weighted density.
While we use the detailed properties of HAE229 to investigate the
impact of environmental processes on the gas properties of galaxies,
Figs.~\ref{fig:tf} and \ref{fig:kslaw} suggest that HAE229, at least
within this context, is not unusual or special.

While it is beyond the scope of this paper to quantitatively constrain
the characteristics of the ICM in high redshift protoclusters, there are
a number of studies of how the ICM may develop.  What we are suggesting
here is that the mass distribution of the (proto)ICM at high redshift
contains a significant fraction of low volume filling factor warm,
$\sim$10$^4$ K, and cold, $\sim$10 K, gas. Accreting gas into massive
halos alone does not explain the entropy profiles of the ICM of local
clusters, even in the absence of cooling \citep[e.g.,][]{voi05}. It
is clear that feedback from active galactic nuclei and star-formation
in cluster galaxies is necessary to provide sufficient entropy and
heating to globally balance the cooling \citep[][]{bes06}. However, the
energy and momentum injection from AGN, such as from the radio galaxy,
MRC\,1138$-$262, is perhaps heating the gas, but also induces substantial
cooling and dissipation \citep[e.g.,][]{emo14,voi15,mee15,emo15b,emo16,gul16}.
Specifically, the large scale environment of MRC\,1138$-$262 has a
substantial Ly$\alpha$ emitting halo \citep{pen97} and a significant mass
of cold molecular gas \citep{emo16}.  Despite likely having relatively
high densities, $\sim0.01-1$~cm$^{-3}$ and $\sim100-1000$~cm$^{-3}$, the
volume filling factors of both these phases are undoubtedly minuscule
($\ll$0.1).  Instead of being a continuous wind of low density hot gas
as in local clusters, these would be more like a intermittent shotgun
blast of cold dense clouds. Under such circumstances the concept of ram
pressure is completely inappropriate. The ISM of the galaxy does not feel
a constant force of the fluid as is assumed in the relation between ram
pressure and the gravitational restoring force \citep{gun72}.  There of
course could still be a diffuse high volume-filling factor hot gas,
but its density must be substantially lower than that in local clusters \citep{car02}.
More work theoretically and observationally is needed to understand our
results \citep{emo16} but the very extended CO disk of HAE229 is already
yielding fascinating clues as to the structure of early intra-cluster
media in protoclusters.

\section{Conclusion}

We have presented the most detailed CO(1-0) observations to date for
a distant, $z_{CO}=2.1478$, normal star-forming galaxy. We detect the
cold molecular gas reservoir of the H$\alpha$ emitter \#229 in very deep
observations with the ATCA. Interestingly, a significant fraction of the
CO emission lies outside the rest-frame UV/optical emitting galaxy. The
physical properties of HAE229 indicates that this source lies on the
main-sequence of galaxies and, its relatively long gas depletion time
and disk morphology suggests that it is the first CO(1-0)-bright HAE
with a ``quiescent mode'' of star-formation.

In order to study environmental dependency of the gas fraction at high
redshifts, we compiled a sample of 24 high-z (proto)cluster members from
the literature. We do not find any environmental dependence suggesting
that usual physical processes seen in local clusters such as harassment,
tidal stripping and ram-pressure stripping do not operate efficiently
at high-redshift in over-dense environments.

In addition, we extend the relation between L$^{\prime}_\mathrm{CO}$  and FWHM
of the CO line previously valid for starbursts \citep{bot13} to gas-rich
main sequence galaxies beyond $z=1$. Our analysis of the integrated
Schmidt-Kennicutt law indicates that the proposed dichotomy between
starbursts and disk galaxies is perhaps more complex than heretofore
known.

Finally, we stress that the number of CO detections of (proto)cluster
galaxies in the distant universe is still very low compared to field
galaxies. In order to understand the influence of the environment on
the molecular gas reservoirs and star-formation efficiency, systematic
surveys of (proto)cluster galaxies beyond $z=1$ in low-order CO lines
must be conducted.

\begin{table*}
\begin{center}
\caption{Positions of HAE229\label{tab:position}.}
\begin{tabular}{lcllcl}
\hline\hline
ID&Instrument&R.A.&Decl.&Offset&Reference\\
&&(J2000.0)&(J2000.0)&$\Delta CO-Other$&\\
\hline
CO counterpart &ATCA&11:40:46.05$\pm$0.05&$-$26:29:11.2$\pm$0.6&---&{\bf this paper}\\
\hline
HAE$\#229$ & VLT-ISAAC&11:40:46.1&$-$26:29:11.5&0.7&\citet{kur04a}\\
HAE$\#902$ & Subaru-MOIRCS&11:40:46.065&$-$26:29:11.33&0.2&\citet{koy13}\\
\hline
\end{tabular}
\tablefoot{Units of right ascension are hours, minutes, and seconds, and units of
declination are degrees, arcminutes, and arcseconds.}
\end{center}
\end{table*}
\begin{table*}
\begin{center}
\caption{Fluxes of HAE229\label{tab:flux}.}
\label{tab:fluxes}
\begin{tabular}{lrccl}
\hline\hline
Band&Unit&HAE229&Instruments&Reference\\
(1)&(2)&(3)&(4)&(5)\\
\hline
$B$&mag&$26.36\pm0.68$&FORS2&\citet{koy13}\\
$Y$&mag&23.62$\pm$0.32&HAWK-I&{\bf this paper}\\
$J$&mag&23.09$\pm$0.03&MOIRCS&\citet{koy13}\\
$H$&mag&$22.26\pm0.10$&HAWK-I&{\bf this paper}\\
$K_{s}$&mag&21.22$\pm$0.08&HAWK-I&{\bf this paper}\\
$K_{s}$&mag&20.86$\pm$0.09&MOIRCS&\citet{doh10}\\
$K_{s}$&mag&21.60$\pm$0.02&MOIRCS&\cite{koy13}\\
$S_{3.6~\mu m}$&mag&19.85$\pm$0.1&IRAC&{\bf this paper}\\
$S_{4.5~\mu m}$&mag&19.83$\pm$0.1&IRAC&{\bf this paper}\\
$S_{5.8~\mu m}$&mag&19.85$\pm$0.2&IRAC&{\bf this paper}\\
$S_{8.0~\mu m}$&mag&20.16$\pm$0.2&IRAC&{\bf this paper}\\
$S_{24.0~\mu m}$&$\mu$Jy&477.4$\pm$5.0&MIPS&\citet{dan14}\\
$S_{100~\mu m}$&mJy&$<$4.5&PACS&\citet{dan14}\\
$S_{160~\mu m}$&mJy&13.6$\pm$4.0&PACS&\citet{dan14}\\
$S_{250~\mu m}$&mJy&26.0$\pm$2.8&SPIRE&\citet{dan14}\\
$S_{350~\mu m}$&mJy&27.2$\pm$2.9&SPIRE&\citet{dan14}\\
$S_{500~\mu m}$&mJy&26.5$\pm$2.7&SPIRE&\citet{dan14}\\
$S_{850~\mu m}$&mJy&2.2$\pm$1.4&SCUBA&\citet{ste03}\\
$S_{CO(1-0)}$&mJy&0.57$\pm$0.06&ATCA&{\bf this paper}\\
$I_{CO(1-0)}$&Jy~km~s$^{-1}$&0.22$\pm$0.03&ATCA&{\bf this paper}\\
\hline
\end{tabular}
\end{center}
\end{table*}
\begin{table*}
\begin{center}
\caption{Properties of HAE229\label{tab:properties}.}
\begin{tabular}{lcl}
\hline\hline
Property&HAE229&Reference\\
\hline
z$_\mathrm{CO(1-0)}$&$2.1478\pm0.0002$&{\bf this paper}\\
z$_\mathrm{H\alpha}$&2.1489&\citet{kur04b}\\
z$_\mathrm{H\alpha}$&2.149&\citet{doh10}\\
FWHM of CO(1-0)&$359\pm34$~km/s&{\bf this paper}\\
FWHM of H$\alpha$&$290\pm60$~km/s&\citet{kur04b}\\
L$^{\prime}_\mathrm{CO(1-0)}$&$5.0\pm0.7~\times~10^{10}$~K~km~s$^{-1}$~pc$^{2}$&{\bf this paper}\\
L$_\mathrm{IR}$&$3.2~\times~10^{12}~L_{\sun}$&{\bf this paper}\\
SFE&$66~L_{\sun}~K~km~s^{-1}~pc^{2}$&{\bf this paper}\\
SFR$_\mathrm{PAH}$&880~M$_{\sun}$ yr$^{-1}$&\citet{ogl12}\\
SFR$_\mathrm{IR}$&555~M$_{\sun}$ yr$^{-1}$&{\bf this paper}\\
SFR$_{H\alpha}$&389~M$_{\sun}$ yr$^{-1}$&\citet{koy13}\\
SFR$_\mathrm{SED}$&$35\pm6$~$~M_{\sun}$ yr$^{-1}$&\citet{doh10}\\
M$_{\star}$&3.7$~\times$~10$^{11}$~M$_{\sun}$&\citet{koy13}\\
M$_{\star}$&5.1$^{+1.5}_{-2.0}~\times$~10$^{11}$~M$_{\sun}$&\citet{doh10}\\
M$_\mathrm{dust}$&$3.5~\times$~10$^{8}$~M$_{\sun}$&{\bf this paper}\\
M$_\mathrm{gas}$&$1.8\pm0.2$~$\times$~10$^{11}$~M$_{\sun}$&{\bf this paper}\\
12 + log(O/H)&8.8&{\bf this paper}\\
$\alpha_\mathrm{CO(1-0)}$&4.0~M$_{\sun}$ pc$^{-2}$~(K~km~s$^{-1}$)$^{-1}$&{\bf this paper}\\
sSFR&1.1~Gyr$^{-1}$&{\bf this paper}\\
M$_\mathrm{gas}$/M$_{\star}$&$0.35-0.49$&{\bf this paper}\\
f$_\mathrm{mol}$&$0.27-0.33$&{\bf this paper}\\
t$_\mathrm{dep}$&0.36~Gyr&{\bf this paper}\\
t$_\mathrm{dyn}$&0.12~Gyr&{\bf this paper}\\
t$_\mathrm{SFR}$&$0.67-0.92$~Gyr&{\bf this paper}\\
\hline
\end{tabular}
\end{center}
\end{table*}

\begin{acknowledgements} 
The Australia Telescope Compact Array is funded by the Commonwealth of Australia for
operation as a National Facility managed by CSIRO.  The authors wish to
express their sincerest thank you to the staff of the CSIRO for their
assistance in conducting these observations and the programme committee
for their generous allocation of time and continuing support for our
research.  H.D. acknowledges financial support from the Spanish Ministry
of Economy and Competitiveness (MINECO) under the 2014 Ram\'{o}n y Cajal
program MINECO RYC-2014-15686.  M.D.L. wishes to thank Gary Mamon for
interesting and entertaining discussions about physical processes that
shape galaxies in clusters. We thank the anonymous referee for her or
his comments that helped us to improve our arguments and presentation in
this paper. B.E. acknowledges funding by the European Union 7th Framework
programme (FP7-PEOPLE-2013-IEF) under grant 624351, and from MINECO grant
AYA2012-32295. N.A.H. acknowledges support from STFC through an Ernest
Rutherford Fellowship and R.O. from CNPq and FAPERJ. Partial support for
D.N. was provided by NSF AST-1009452, AST-1445357, NASA \textit{HST} AR-13906.001
from the Space Telescope Science Institute, which is operated by the
Association of University for Research in Astronomy, Incorporated,
under NASA Contract NAS5-26555, and a Cottrell College Science Award,
awarded by the Research Corporation for Science Advancement. Our results
are partially based on observations made with ESO Telescope at Paranal
under programmes 088.A-074(B), 091.A-0106(A) and 094.A-0104(A) and with
the Hubble Space telescope.

\end{acknowledgements}

\begin{appendix}

\section{Properties of distant galaxies in overdensities}

\clearpage

\begin{landscape}
\begin{table}
\footnotesize
\begin{center}
\caption{CO observations of $z>0.4$ cluster members\label{tab:cocluster}.}
\label{clusterco}
\begin{tabular}{lcccccccc}
\hline\hline
Name&$z_{CO}$&Transition&I$_\mathrm{CO}$&FWHM&Telescope&L$^{\prime}_\mathrm{CO}$&L$_\mathrm{IR}$&Reference\\
&&&(Jy~km$^{\prime}$)&(km~s$^{-1}$&&($10^{10}$~K~km~s$^{-1}$~pc$^{2}$)&($10^{12}$)&\\
\hline
\multicolumn{2}{l}{\it Cluster Cl0024$+$16 at $z=0.40$}\\
MIPS~J002652.5&0.3799&$1-0$&&$140\pm10$&PdBI&$0.64\pm0.05$&$3.5\pm0.5$&\citet{gea11}\\
MIPS~J002621.7&0.3803&$1-0$&&$144\pm14$&PdBI&$0.68\pm0.06$&$3.1\pm0.2$&\citet{gea09}\\
MIPS~J002715.0&0.3813&$1-0$&&$340\pm40$&PdBI&$0.26\pm0.03$&$1.9\pm0.3$&\citet{gea11}\\
MIPS~J002703.6&0.3956&$1-0$&&$250\pm30$&PdBI&$0.43\pm0.06$&$2.3\pm0.3$&\citet{gea11}\\
MIPS~J002721.0&0.3964&$1-0$&&$158\pm34$&PdBI&$1.14\pm0.11$&$3.2\pm0.2$&\citet{gea09}\\
\hline
\multicolumn{2}{l}{\it Cluster Cl1416$+$4446 at $z=0.40$}\\
GAL1416$+$446&0.3964&$1-0$&$1.0\pm0.1$&$420\pm40$&PdBI&$0.81\pm0.8$&$0.275$&\citet{jab13}\\
\hline
\multicolumn{2}{l}{\it Cluster Cl09266$+$1242 at $z=0.49$}\\
GAL0926$+$1242$-$A&0.4886&$2-1$&$0.6\pm0.1$&$420\pm40$&PdBI&$0.19\pm0.03$&$0.165$&\citet{jab13}\\
GAL0926$+$1242$-$B&0.4886&$2-1$&$0.5\pm0.1$&$200\pm20$&PdBI&$0.16\pm0.03$&$0.082$&\citet{jab13}\\
\hline
\multicolumn{2}{l}{\it Cluster 7C~1756$+$6520 at $z=1.42$}\\
AGN.1317&$1.4161\pm0.0001$&$2-1$&$0.52\pm0.06$&$254\pm33$&PdBI&$1.36\pm0.15$&&\cite{cas13}\\
\hline
\multicolumn{2}{l}{\it Cluster COSMOS at $z=1.55$}\\
51613&$1.517$&$1-0$&$0.20\pm0.05$&$200\pm80$&VLA&$2.42\pm0.58$&$0.57$&\citet{ara12}\\
51858&$1.556$&$1-0$&$0.10\pm0.03$&$360\pm220$&VLA&$1.26\pm0.38$&$0.95$&\citet{ara12}\\
\hline
\multicolumn{2}{l}{\it protocluster MRC\,1138$-$262 at $z=2.16$}\\
HAE229&$2.1480\pm0.0004$&$1-0$&$0.22\pm0.02$&$359\pm34$&ATCA&$5.0\pm0.7$&$3.2$&{\bf this paper}\\
\hline
\multicolumn{2}{l}{\it protocluster HATLAS~J084933 at $z=2.41$}\\
HATLAS~J084933 W&$2.4066\pm0.0006$&$1-0$&$0.49\pm0.06$&$825\pm115$&VLA&$13.8\pm1.7$&$33.1^{+3.2}_{-2.9}$&\citet{ivi13}\\
HATLAS~J084933 T&$2.4090\pm0.0003$&$1-0$&$0.56\pm0.07$&$610\pm55$&VLA&$15.7\pm2.0$&$14.5^{+1.8}_{-1.6}$&\citet{ivi13}\\
HATLAS~J084933 M&$2.4176\pm0.0004$&$1-0$&$0.057\pm0.013$&$320\pm70$&VLA&$1.6\pm0.4$&$7.9^{+4.6}_{-2.9}$&\citet{ivi13}\\
HATLAS~J084933 C&$2.4138\pm0.0003$&$1-0$&$0.079\pm0.014$&$250\pm100$&VLA&$2.2\pm0.4$&$6.3^{+3.7}_{-2.3}$&\citet{ivi13}\\
\hline
\multicolumn{2}{l}{\it protocluster USS~1558$-$003 at $z=2.51$}\\
rHAE$-$193&$2.5131$&$1-0$&$0.096\pm0.015$&$437$&VLA&$2.8$&$5.1$&\citet{tad14}\\
bHAE$-$191&$2.5168$&$1-0$&$0.052\pm 0.008$&$251$&VLA&$1.5$&$2.5$&\citet{tad14}\\
\hline
\multicolumn{2}{l}{\it protocluster B3~J2330 at $z=3.09$}\\
JVLA~J233024.69$+$392708.6&$3.0884\pm0.0010$&$1-0$&$0.16\pm0.03$&$720\pm170$&VLA&$6.9\pm1.5$&&\citet{ivi12}\\
\hline
\multicolumn{2}{l}{\it protocluster GN20 at $z=4.05$}\\
GN20&$4.0548\pm0.0008$&$2-1$&$1.0\pm0.3$&$310\pm110$&VLA&$16.0\pm5.0$&$18.6^{+0.9}_{-0.8}$&\citet{hod12,tan14}\\
GN20.2a&$4.051\pm0.001$&$2-1$&$0.6\pm0.2$&$830\pm190$&VLA&$9.7\pm2.9$&$7.9 ^{+0.4}_{-0.9} $&\citet{hod13a,tan14}\\
GN20.2b&$4.056\pm0.001$&$2-1$&$0.3\pm0.2$&$400\pm210$&VLA&$4.2\pm2.9$&$6.9 ^{+0.7}_{-1.4} $&\citet{hod13a,tan14}\\
\hline
\multicolumn{2}{l}{\it protocluster HDF850.1 at $z=5.18$}\\
HDF850.1&$5.183$&$2-1$&$0.17\pm0.04$&$400\pm30$&VLA&$4.1$&$8.7\pm1.0$&\citet{wal12}\\
\hline
\multicolumn{2}{l}{\it protocluster AzTEC-3 at $z=5.30$}\\
AzTEC$-$3&$5.2979\pm 0.0004$&$2-1$&$0.23\pm0.03$&$487\pm58$&VLA&$5.84\pm0.78$&$17\pm8$&\citet{rie10}\\
\hline
\end{tabular}
\end{center}
\end{table}
\end{landscape}

\end{appendix}

\end{document}